\documentclass[aps,prd,reprint,twocolumn,nofootinbib,longbibliography,floatfix,preprintnumbers]{revtex4-2}
\usepackage[T1]{fontenc}
\usepackage{amsmath,amssymb,mathrsfs,bm,booktabs}
\usepackage{graphicx}
\graphicspath{{figures/}{../figures/}}
\usepackage[hidelinks]{hyperref}
\hypersetup{pdftitle={Horizon-penetrating rotating black-hole configurations in VCDM},pdfauthor={Antonio De Felice}}

\newcommand{\dd}{\mathrm{d}}
\newcommand{\Le}{\Lambda_{\mathrm{eff}}}
\newcommand{\HL}{H_{\Lambda}}
\newcommand{\Dr}{\Delta_r}
\newcommand{\Dz}{\Delta_z}
\newcommand{\IR}{\mathscr I^+}
\newcommand{\QQ}{\mathcal Q}
\newcommand{\WW}{\mathcal W}
\newcommand{\sigmasq}{\sigma_{ij}\sigma^{ij}}

\begin{document}
\preprint{YITP-26-156}

\title{Horizon-penetrating rotating black-hole configurations in VCDM}
\author{Antonio De Felice}
\affiliation{Center for Gravitational Physics and Quantum Information,
Yukawa Institute for Theoretical Physics, Kyoto University, Kyoto
606-8502, Japan}
\date{November 9, 2026}

\begin{abstract}
A regular black-hole metric need not define a regular configuration of
VCDM, whose preferred foliation is part of the gravitational solution.
We construct stationary rotating, possibly charged configurations on the
constant-auxiliary Einstein--Maxwell branch. Their Kerr--Newman and
Kerr--Newman--de Sitter geometries carry nonzero constant-mean-curvature
foliations that cross the future black-hole horizon and, in the de Sitter
case, the future cosmological horizon. A horizon-regular flux formulation
specifies the inner data and the distinct hyperboloidal and expanding
asymptotic ends. Seventeen survey configurations are supplemented by a
mean-flux scan at fixed four-geometry and paired flat/de Sitter comparisons
down to $\kappa M=0.01$, where $K_0=3\kappa$. Varying the flux changes the
preferred-foliation shear without changing the background seen by standard
test matter. The paired local lapse and shear differences have effective
powers close to two over the computed range, with the lapse comparison
requiring a fixed time normalization. Continuous residuals, geometric
reconstructions, and selected grid comparisons test the numerical
solutions. An exact lapse bound excludes zeros in the regular stationary
exterior block with $\Dr>0$. 
We distinguish singular finite-proper-distance endpoints from asymptotic
trumpet ends and give conditional bounds on a stationary spacelike inner
limit. A local cosmological interpretation requires additional matching
assumptions; dynamical selection and the rotating interior endpoint are
not determined by the stationary construction.
\end{abstract}

\maketitle

\section{Introduction}\label{sec:intro}
A theory motivated by cosmological expansion must also admit regular
local strong-field configurations. In VCDM, identifying a familiar
black-hole four-metric is not sufficient: the gravitational configuration
also specifies a preferred foliation and auxiliary fields. The theory
has the two local propagating tensor degrees of freedom of general
relativity (GR), but a different constraint structure
\cite{DeFeliceDollMukohyama2020,DeFeliceMaedaMukohyamaPookkillath2022}. Regularity of the preferred
normal must therefore be examined together with regularity of the metric.

The known embedding of GR solutions into VCDM illustrates this distinction.
Ref.~\cite{DeFeliceMaedaMukohyamaPookkillath2022} established the
constant-auxiliary embedding criterion and a maximal, $K_0=0$,
Boyer--Lindquist (BL) realization of Kerr--de Sitter. When the preferred
leaves are the constant-BL-time surfaces, their normal acceleration
diverges at a simple metric horizon, although the four-metric extends
smoothly through it. This foliation does not give a regular
future-horizon crossing. The obstruction is specific to the BL
foliation, not to every maximal slicing. A divergent gradient of a time
label does not alone classify an endpoint. However, a lapse tending to
zero along a leaf while remaining finite and positive on another portion
of that leaf cannot be regularized by a time relabeling on a region
covering both portions. Section~\ref{sec:clockregularity} distinguishes
this obstruction from local singularity and gives a finite-proper-distance
criterion in terms of the normal acceleration.

Here we construct preferred foliations of Kerr--Newman and
Kerr--Newman--de Sitter with nonzero constant mean curvature (CMC),
$K=K_0$. Their domains extend from an inner surface inside the outer
black-hole horizon to a hyperboloidal or expanding de Sitter end. The
preferred normal crosses the future black-hole horizon and, on the dS
branch, the future cosmological horizon. We start with the GR metric,
solve for a stationary height function $t=\tau+h(r,z)$, and adopt
$T=t$ as the preferred time, subject to the VCDM auxiliary equations.
This specifies a new configuration; it is not a passive coordinate
change of an already assigned clock $T_{\rm old}=\tau$.

The construction builds on established geometric results. Nonzero-CMC
slices of Kerr and freedom in their inner angular data were studied by
Schinkel, Macedo, and Ansorg~\cite{SchinkelMacedoAnsorg2014}; maximal
Kerr--Newman slices by Duncan~\cite{Duncan1985}. Rotating trumpet
coordinates, including charged de Sitter extensions, need not be CMC
\cite{DennisonBaumgarteMontero2014}. Spherical Schwarzschild--de Sitter
CMC slices and de Sitter trumpet constructions are also known
\cite{NakaoMaedaNakamuraOohara1991,DennisonBaumgarte2017}, as are
hyperboloidal Kerr--Newman foliations~\cite{Winicour2026}. Horizon
penetration or hyperboloidal asymptotics alone does not prescribe the
mean curvature throughout a leaf~\cite{Zenginoglu2025}.

Our contribution is the explicit completion of rotating charged
Einstein--Maxwell geometries into horizon-penetrating VCDM configurations
with compatible auxiliary fields and specified inner and asymptotic
data. In particular, the charged dS construction combines nonzero CMC,
rotation, and crossing of both future horizons. It is not a new
Einstein--Maxwell four-metric or a new general embedding theorem. The
flux formulation, asymptotic coefficients, regularity criteria, and
controlled numerical realizations provide the preferred-time data that
those earlier results do not by themselves specify.

An original seventeen-configuration survey is supplemented by two scans.
At fixed four-geometry and $K_0$, varying the mean normal flux $C$ changes
the scalar shear of the preferred foliation. Changing its zero-mean
inner angular profile gives a further distinction at fixed $C$. Paired
AF/dS calculations at $\kappa M=0.10,0.05,0.03,0.01$ test local similarity
as the mean-curvature scale decreases. The comparison fixes the displayed
metric parameters, coordinate identification, and time normalization;
it is not an evolution in one specified cosmological potential.

These stationary data do not select their own formation history.
Spherical collapse studies show the relevance of the preferred interior
slicing but do not determine its rotating counterpart
\cite{DeFeliceDollLarrouturouMukohyama2021,DeFeliceMaedaMukohyamaPookkillath2023,JalaliMartensMukohyama2024}.
We distinguish the solved boundary-value problem from interior
continuation, collapse selection, and matching to matter-filled FLRW.
The cosmological discussion supplies explicit conditions under which a
stationary configuration can be a local reference, rather than claiming
that matching has already been constructed.

Section~\ref{sec:vcdm} specifies the VCDM branch.
Sections~\ref{sec:pde} and \ref{sec:zero} develop the regular formulation
and endpoint criteria. Sections~\ref{sec:asympt}--\ref{sec:results}
present the asymptotic data, numerical method, and configurations.
Section~\ref{sec:discussion} discusses matter probes and the conditional
cosmological interpretation. The appendices record coefficients,
benchmarks, and the scope of numerical validation.

We use signature $(-+++)$, $G=c=1$, and
$K_{\mu\nu}=\gamma_\mu{}^\alpha\gamma_\nu{}^\beta\nabla_\alpha n_\beta$,
so expanding flat de Sitter slices have $K>0$. The geometric mass
parameter $M$, spin parameter $a$, and charge parameter $q$ have units
of length, $\kappa=K_0/3$ of inverse length, and $C$ of length squared.
$T=t$ in the adapted chart, whereas $\tau=t_{\rm BL}$ denotes BL time.
Metric parameters are not automatically identified with de Sitter
conserved charges; all comparisons use the stated normalization.

\section{The Einstein--Maxwell branch of VCDM}\label{sec:vcdm}
\subsection{Action, 3+1 decomposition, and auxiliary conditions}
For a time function $T$ with timelike gradient, define the
future-directed normal and induced geometry by
\begin{align}
 n_\mu&=-N\partial_\mu T,&
 N^{-2}&=-g^{\mu\nu}\partial_\mu T\partial_\nu T,\\
 \gamma_{\mu\nu}&=g_{\mu\nu}+n_\mu n_\nu,&
 K_{\mu\nu}&=\gamma_\mu{}^\alpha\gamma_\nu{}^\beta\nabla_\alpha n_\beta.
\end{align}
Here $N>0$ and $K=\nabla_\mu n^\mu$. In unitary gauge $T=t$, the
variables are the lapse, shift, spatial metric, and auxiliaries
$\varphi$, $\lambda$, and $\lambda_2$. With a minimally coupled,
geometrically normalized Maxwell field
$F_{\mu\nu}=\partial_\mu\mathsf A_\nu-\partial_\nu\mathsf A_\mu$, the
bulk action is~\cite{DeFeliceDollMukohyama2020,DeFeliceMaedaMukohyamaPookkillath2022}
\begin{align}\label{eq:action}
 S={1\over16\pi}\int&\dd t\,\dd^3x\,N\sqrt\gamma\,\mathcal L,\nonumber\\
 \mathcal L={}&{}^{(3)}R+K_{ij}K^{ij}-K^2-2V(\varphi)\nonumber\\
 &-\frac32\lambda^2-2\lambda(K+\varphi)
 +\frac{2\lambda_2}{N}D^2\varphi\nonumber\\
 &-F_{\mu\nu}F^{\mu\nu}.
\end{align}
$D_i$ is the spatial covariant derivative and $D^2=D_iD^i$.
The first three terms are the ADM bulk form of the Einstein--Hilbert
action. Suitable surface terms and fixed boundary data are understood.

Varying the auxiliaries \emph{before} restricting the solution gives
\begin{align}\label{eq:aux}
 \lambda&=-\frac23(K+\varphi),\\
 D^2\varphi&=0,&D^2\lambda_2&=N[V'(\varphi)+\lambda].
\end{align}
In particular, setting $\lambda_2=0$ does not remove its companion
elliptic equation: its source must vanish. We take the sufficient branch
\begin{equation}\label{eq:branch}
 K=K_0,\qquad\varphi=\varphi_0,\qquad
 \lambda=\lambda_0,\qquad\lambda_2=0,
\end{equation}
with spacetime-constant $K_0,\varphi_0,\lambda_0$ and compatible inner
and asymptotic auxiliary data. Equations~\eqref{eq:aux} then require
\begin{align}\label{eq:potential}
 \lambda_0&=-\frac23(K_0+\varphi_0),&V'(\varphi_0)&=-\lambda_0,\\
 \Le&=V(\varphi_0)+\frac{K_0^2-\varphi_0^2}{3}.
 \label{eq:lambdaeff}
\end{align}
The last relation follows from the metric equations. Indeed, terms
containing $\lambda_2$ vanish on this branch, while
\[
 \sqrt{-g}\,\lambda_0 K
 =\lambda_0\partial_\mu(\sqrt{-g}\,n^\mu)
\]
is a total divergence. Its bulk variation is proportional to derivatives
of $\lambda$ before the branch is imposed. The remaining algebraic terms
satisfy
\[
 2V(\varphi_0)+\frac32\lambda_0^2+2\lambda_0\varphi_0
 =2\left[V(\varphi_0)+\frac{K_0^2-\varphi_0^2}{3}\right].
\]
The metric and Maxwell equations consequently reduce to
\begin{align}\label{eq:EM}
 G_{\mu\nu}+\Le g_{\mu\nu}
 &=2\left(F_{\mu\alpha}F_\nu{}^\alpha-\tfrac14g_{\mu\nu}F^2\right),\\
 \nabla_\mu F^{\mu\nu}&=0.
\end{align}
This is the constant-auxiliary GR embedding, here applied to electrovacuum.

\paragraph{Closure of the preferred-time equation.}
The covariant formulation also contains an equation for $T$; working in
unitary gauge must not discard it. In the covariant equations of
Ref.~\cite{DeFeliceMaedaMukohyamaPookkillath2022}, the multiplier enforcing
unit-normal normalization is zero when $\nabla_\mu\lambda=0$,
$\nabla_\mu\varphi=0$, and $\lambda_2=0$. The remaining terms in the
$T$ equation contain these derivatives or $\lambda_2$ and vanish.
Equivalently, after imposing the independent auxiliary equations, the
only surviving $T$-dependent bulk term is $\lambda_0\nabla_\mu n^\mu$,
whose variation is a divergence for compactly supported variations.
Thus no additional bulk clock equation is left unsatisfied on this
branch. This argument does not replace the boundary data for the
elliptic foliation problem.

\paragraph{A compatible local potential.}
The branch fixes the value and first derivative of $V$, not its whole
function. For example, with $\varphi_0=0$ as part of the model
specification,
\begin{equation}\label{eq:localpotentialexample}
 V(\varphi)=\Le-3\kappa^2+2\kappa\varphi
            +\frac{\nu}{2}\varphi^2
\end{equation}
satisfies Eqs.~\eqref{eq:potential}--\eqref{eq:lambdaeff} with
$K_0=3\kappa$ and $\lambda_0=-2\kappa$. The dimensionless curvature
$\nu$ is not fixed by this background. This example establishes local
compatibility, not a realistic cosmology or stability of the chosen
potential. In general we retain the algebraic branch conditions rather
than selecting one potential for the whole parameter survey.

\subsection{Asymptotic state and data characterizing the solution}
\label{sec:asymptoticstate}\label{sec:hair}
Two scales must be distinguished:
\begin{equation}\label{eq:scales}
 \kappa=K_0/3>0,\qquad \HL^2=\Le/3\ge0.
\end{equation}
The cosmological family computed here has $\kappa=\HL$; the
asymptotically flat family has $\HL=0$ but $\kappa>0$.
In the latter, the four-metric is asymptotically flat while the
preferred leaves are hyperboloidal, not asymptotically Euclidean.
The code variable \texttt{H} denotes $\kappa$ in both cases.

Eliminating $\lambda_0$ from the auxiliary conditions gives
\begin{align}\label{eq:K0fixed}
 K_0&=\frac32V'(\varphi_0)-\varphi_0,\\
 \Le&=V(\varphi_0)-\varphi_0V'(\varphi_0)
                    +\frac34[V'(\varphi_0)]^2.
 \label{eq:fixedpointLambda}
\end{align}
These relations connect the stationary state to the chosen potential
and auxiliary value. Since $K$ is constant,
\begin{equation}\label{eq:Kboundary}
 K_0=K_\infty.
\end{equation}
On the matched de Sitter branch, $K_\infty=3H_\Lambda$ and
$V(\varphi_0)=\varphi_0^2/3$. On the flat branch, $K_\infty=3\kappa$
sets the asymptotic opening of the hyperboloidal leaves, despite
$\Le=0$. Varying $M,a,q$ or the foliation boundary data does not
change the algebraic conditions, although regularity restricts which
values are admissible. Changing $K_0$ or $\Le$ may require a different
compatible auxiliary state or potential. In particular, the auxiliary
relations do not guarantee that a prescribed potential admits both
members of an AF/dS pair at the same $K_0$. Our paired scan compares
compatible stationary boundary-value configurations; it is not an
independent variation of $\Le$ within one fixed auxiliary state.
Section~\ref{sec:cosmomatch}
explains the instantaneous connection to a fixed evolving cosmology.

The remaining physical specification is not exhausted by the metric
parameters and $K_\infty$. The flux first integral below introduces
$C$, while the zero-mean angular part $\delta P$ of the inner flux
provides additional boundary data. For a chosen inner surface and
outer prescription, it is these data that distinguish the stationary
configurations examined here. In a formation problem they should be
outputs of the evolution and matching to the interior, rather than
freely reset boundary choices. We therefore focus on their geometric
and dynamical roles, without requiring a classification as black-hole
hair. The observability of the resulting foliation differences is
addressed in Sec.~\ref{sec:matterprobes}.

\section{Stationary CMC geometry and regular horizon crossing}\label{sec:pde}
\subsection{Background metric and electromagnetic field}
Having established that the four-dimensional metric and Maxwell field on the constant-auxiliary branch satisfy the standard Einstein--Maxwell--$\Le$ equations~\eqref{eq:EM}, we now specify the stationary, axisymmetric black-hole background explicitly~\cite{Carter1968}. Instead of the polar angle $\theta \in [0,\pi]$, it is convenient to work with the directional cosine $z = \cos\theta \in [-1, 1]$, so that $\sin^2\theta = 1 - z^2$ and $\dd\theta^2 = \dd z^2 / (1 - z^2)$. We define the standard Kerr--Newman--de Sitter metric functions
\begin{align}
 J&=r^2+a^2,&\Sigma&=r^2+a^2z^2,\\
 \Dr&=J(1-\HL^2r^2)-2Mr+q^2,\\
 \Dz&=1+\HL^2a^2z^2,&\Xi&=1+\HL^2a^2.
\end{align}
Here $M$ is the mass parameter, $a$ is the specific angular momentum, $q$ is the electric charge parameter, and $\HL = \sqrt{\Le/3}$ is the cosmological inverse-length scale (with $\HL = 0$, $\Dz = \Xi = 1$ in the asymptotically flat Kerr--Newman limit).

In Boyer--Lindquist-type coordinates $(\tau, r, z, \phi)$, using the Killing-time and azimuthal normalization adopted in our numerical code, the background line element reads
\begin{align}\label{eq:BL}
 \dd s^2={}&-\frac{\Dr}{\Xi^2\Sigma}
 [\dd\tau-a(1-z^2)\dd\phi]^2\nonumber\\
 &+\frac{\Sigma}{\Dr}\dd r^2
 +\frac{\Sigma}{(1-z^2)\Dz}\dd z^2\nonumber\\
 &+\frac{(1-z^2)\Dz}{\Xi^2\Sigma}
 [a\dd\tau-J\dd\phi]^2.
\end{align}
The azimuthal normalization gives a regular axis for $2\pi$ periodicity
of $\phi$; the time normalization is the convention used by the solver.
The coordinate volume element is
\[
 \sqrt{-g}=\frac{\Sigma}{\Xi^2}.
\]

The background Maxwell field $F_{\mu\nu}$ is generated by the four-potential one-form
\begin{equation}\label{eq:EMpotential}
 \mathsf A=-{qr\over\Xi\Sigma}
 [\dd\tau-a(1-z^2)\dd\phi],\qquad F=\dd\mathsf A.
\end{equation}
We refer to $\mathsf A$ in Eq.~\eqref{eq:EMpotential} as a gauge representative because the physical electromagnetic field $F = \dd\mathsf A$ is invariant under $U(1)$ gauge transformations $\mathsf A \mapsto \mathsf A + \dd\chi$. Physical regularity of the electromagnetic field across a horizon is a statement about the curvature two-form $F$; when changing from Boyer--Lindquist coordinates to horizon-penetrating coordinates, one is always free to perform a gauge shift $\dd\chi(r)$ so that the one-form $\mathsf A$ itself also remains finite at $\Dr = 0$.

\subsection{Why the Boyer--Lindquist time is not sufficient}\label{sec:BLtime}
As a separate configuration, choose $T_{\rm BL}=\tau$. In BL coordinates
$g^{\tau r}=g^{\tau z}=0$, so its normal has only stationary $\tau$ and
$\phi$ components. Hence
\begin{equation}\label{eq:BLK}
 K_{\rm BL}=\frac{1}{\sqrt{-g}}\partial_\mu
               (\sqrt{-g}\,n^\mu)=0.
\end{equation}
The leaves are maximal, as in
Ref.~\cite{DeFeliceMaedaMukohyamaPookkillath2022}; this fixes the trace,
not the full tensor $K_{ij}$. Where the BL leaves are spacelike,
\begin{equation}\label{eq:BLlapse}
 N_{\rm BL}^2=\frac{\Sigma\Dr}{\WW},\qquad
 \WW=\Xi^2\left[J^2-\frac{a^2(1-z^2)\Dr}{\Dz}\right].
\end{equation}
At a simple horizon $r_h$, $\Dr$ has a simple zero, whereas
$\WW(r_h,z)=\Xi^2(r_h^2+a^2)^2$ and $\Sigma$ remain positive and finite.
Consequently,
\begin{align}\label{eq:BLclocknorm}
 X_{T_{\rm BL}}&\equiv-g^{\mu\nu}\nabla_\mu T_{\rm BL}
                         \nabla_\nu T_{\rm BL}\nonumber\\
 &=N_{\rm BL}^{-2}=\frac{\WW}{\Sigma\Dr}\longrightarrow+\infty
 \qquad(\Dr\to0^+).
\end{align}
This obstructs a regular extension of the prescribed BL clock. A
stronger, label-independent obstruction follows from the normal's
four-acceleration. With $a_i=D_i\ln N_{\rm BL}$,
$\gamma^{rr}=\Dr/\Sigma$, and
$\ln N_{\rm BL}=\tfrac12\ln\Dr+O(1)$,
\begin{equation}\label{eq:BLacceleration}
 a_\mu a^\mu=a_i a^i
 =\frac{[\Dr'(r_h)]^2}{4\Sigma(r_h,z)\Dr}+O(1).
\end{equation}
Thus the preferred normal itself has no smooth extension through a simple
horizon, although the four-metric does. This is not a spacetime curvature
singularity or a prediction of a matter ``firewall''.

A passive coordinate change preserves an assigned scalar
$T_{\rm old}=\tau$ and its normal; it cannot cure this obstruction.
We instead choose different CMC leaves and compatible auxiliary data.
The distinction between a singular time label and a singular normalized
foliation, which is immaterial to the conclusion from
Eq.~\eqref{eq:BLacceleration} but essential for general lapse zeros,
is developed in Sec.~\ref{sec:clockregularity}.

\subsection{The preferred time and normal fluxes}\label{sec:height}
Starting from the GR metric, introduce a stationary axisymmetric height
function and then assign the preferred clock,
\begin{equation}\label{eq:heightfunction}
 t=\tau+h(r,z),\qquad T\equiv t.
\end{equation}
A leaf satisfies $\tau=t_*-h(r,z)$. Since $\partial_\tau=\partial_t$
at fixed spatial coordinates, stationarity is preserved. By contrast,
a passive transformation of the old assignment $T_{\rm old}=\tau$
would give $T_{\rm old}=t-h$ and leave its maximal foliation unchanged.

The convention $\partial_\tau t=1$ fixes all quoted lapses. It does not
identify $t$ with cosmological proper time: in the dS reference
$t_{\rm cos}=t/\Xi$, whereas the AF hyperboloidal end has no globally
comoving proper-time normalization. We keep this distinction in the
paired comparison and in Sec.~\ref{sec:cosmomatch}.

We restrict to the reflection-symmetric subclass $h(r,-z)=h(r,z)$ and
solve on $0\le z\le1$. At the equator $h_z=0$. Axis regularity requires
smoothness in local Cartesian coordinates: a smooth axisymmetric
function of $z$ has finite $h_z$ and hence
$h_\theta=-\sin\theta\,h_z\to0$ on the axis. It does not require
$h_z(1)=0$, or imply that every admissible foliation shares reflection
symmetry.

Set $b=(1-z^2)\Dz$ and
\begin{equation}
 \mathcal A=\Xi^2\left(\frac{J^2}{\Dr}
             -\frac{a^2(1-z^2)}{\Dz}\right)=\frac{\WW}{\Dr}.
\end{equation}
Then $g^{\tau\tau}=-\mathcal A/\Sigma$,
$g^{rr}=\Dr/\Sigma$, and $g^{zz}=b/\Sigma$. With
$\partial_\mu T=(1,h_r,h_z,0)$,
\begin{equation}\label{eq:inverseTT}
 g^{tt}=\frac{-\mathcal A+\Dr h_r^2+b h_z^2}{\Sigma},
\end{equation}
and a spacelike leaf has
\begin{equation}\label{eq:lapse}
 N^2=\frac{\Sigma}{\mathcal A-\Dr h_r^2-bh_z^2}>0.
\end{equation}
It is this transformed component, not $g^{\tau\tau}$, that determines
the preferred lapse.

\paragraph{Tilt relative to BL.}
Where the two normals are future-directed and timelike, define
$-n_\mu n_{\rm BL}^\mu=\cosh\eta$. Their common stationary-time
normalization gives
\begin{align}\label{eq:BLboost}
 \cosh\eta&=\frac{N}{N_{\rm BL}},&
 \tanh^2\eta&=1-\frac{N_{\rm BL}^2}{N^2}
 =\frac{\Dr h_r^2+b h_z^2}{\mathcal A}.
\end{align}
An absolute lapse difference alone does not measure tilt. For
$N/N_{\rm BL}-1\ll1$, $\eta\simeq[2(N/N_{\rm BL}-1)]^{1/2}$.
For the $a/M=0.6$, $q/M=0.4$, $\kappa M=0.1$, $C/M^2=2$ benchmarks at
$(r/M,z)=(3,0.5)$, the preferred lapses are $0.613796328185$ (AF) and
$0.534384311671$ (dS), versus $N_{\rm BL}=0.607639413110$ and
$0.527370694200$. Thus $\tanh\eta\simeq0.1413$ and $0.1615$.
At a regularly crossed horizon $N$ stays positive while
$N_{\rm BL}\to0$; the divergent relative rapidity reflects the
breakdown of the BL reference normal, not of the regular preferred
normal. The comparison is not defined where BL leaves cease to be
spacelike.

The meridional normal components motivate oriented fluxes
\begin{equation}\label{eq:fluxes}
 P=N\Dr h_r=-\Sigma n^r,\qquad
 \QQ=Nb h_z=-\Sigma n^z.
\end{equation}
Using $\sqrt{-g}=\Sigma/\Xi^2$, stationarity and axisymmetry give
\begin{equation}\label{eq:Kflux}
 K=-\frac{1}{\Sigma}(\partial_rP+\partial_z\QQ),
\end{equation}
so the exact CMC equation is
\begin{equation}\label{eq:CMC}
 \partial_rP+\partial_z\QQ=-3\kappa\Sigma.
\end{equation}
Reflection symmetry gives $\QQ(r,0)=0$, while axis smoothness gives
$\QQ=O(1-z^2)$ at $z=1$. Angular integration and then radial integration
yield
\begin{equation}\label{eq:firstint}
 \int_0^1P(r,z)\dd z=C-\kappa r(r^2+a^2).
\end{equation}
The constant $C$ fixes the shifted angular-mean flux, not the pointwise
angular profile. Inner angular data are an additional part of the
boundary-value specification.

\subsection{A regular radial reference and horizon pole cancellation}
For a smooth spacelike crossing of a simple horizon, the lapse--flux
identity derived below gives
\begin{equation}
 P_h^2=\Xi^2(r_h^2+a^2)^2N_h^2,
 \qquad
 \lim_{r\to r_h}\Dr h_r=\pm\Xi(r_h^2+a^2).
\end{equation}
The sign is positive for the future black-hole horizon and negative for
the future cosmological horizon. The pole coefficient is independent of
angle, although $P_h$ and $N_h$ need not be. The first integral fixes
only the angular mean of $P_h$, not its pointwise sign.

Both $h_r$ and $\mathcal A$ contain BL-coordinate poles. Their presence
alone does not signal a singular foliation.\footnote{For Schwarzschild,
the ingoing Painlev\'e--Gullstrand time satisfies
$\dd T_{\rm PG}=\dd\tau+\sqrt{2M/r}\,(1-2M/r)^{-1}\dd r$.
Its height derivative diverges at $r=2M$, while the transformed metric
and normal are regular and $N_{\rm PG}=1$. This illustrates pole
cancellation, not a CMC example.}
We cancel the poles analytically by writing
$h=h_{\rm ref}(r)+u(r,z)$ and defining
\begin{align}\label{eq:radialreference}
 \bar P&=C-\kappa rJ,&\mathcal F&=\bar P^2+\Dr J,\\
 s&=\bar P/\sqrt{\mathcal F},&h_r&=\Xi Js/\Dr+u_r,\\
 &&h_z&=u_z.\nonumber
\end{align}
Since $1-s^2=\Dr J/\mathcal F$,
\begin{equation}\label{eq:polecancellation}
 \frac{\Xi^2J^2}{\Dr}(1-s^2)=\frac{\Xi^2J^3}{\mathcal F}.
\end{equation}
The lapse and fluxes consequently have the horizon-regular forms
\begin{align}\label{eq:D}
 N^2&=\Sigma/\mathcal D,\\
 \mathcal D&=\frac{\Xi^2J^3}{\mathcal F}
 -\frac{\Xi^2a^2(1-z^2)}{\Dz}
 -2\Xi Js u_r-\Dr u_r^2-bu_z^2,\nonumber
\end{align}
\[
 P=N(\Xi Js+\Dr u_r),\qquad \QQ=Nb u_z.
\]
Our reference chart requires $\mathcal F>0$, and spacelike preferred
leaves require $\mathcal D>0$. These are distinct conditions:
$\mathcal F$ depends only on the chosen radial reference, whereas
$\mathcal D$ depends on the solved derivatives. In the spherical
subfamily $u=0$ and $N^2=\mathcal F/r^4$; for rotation, a zero of
$\mathcal F$ instead marks failure of this decomposition and does not
by itself locate or exclude a physical lapse zero.

To regularize the four-metric, accompany
$\dd\tau=\dd t-\dd u-(\Xi Js/\Dr)\dd r$ by
\begin{equation}\label{eq:azimuthshift}
 \dd\phi=\dd\psi-\frac{\Xi a s}{\Dr}\dd r.
\end{equation}
With $e=\dd t-\dd u-a(1-z^2)\dd\psi$ and
$f=a(\dd t-\dd u)-J\dd\psi$, the identity
$J-a^2(1-z^2)=\Sigma$ gives
\begin{align}\label{eq:coframecancellation}
 \dd\tau-a(1-z^2)\dd\phi&=e-\frac{\Xi s\Sigma}{\Dr}\dd r,\\
 a\dd\tau-J\dd\phi&=f.
\end{align}
The cancelling radial poles leave
\begin{align}\label{eq:regularmetric}
 \dd s^2={}&-\frac{\Dr}{\Xi^2\Sigma}e^2
 +\frac{2s}{\Xi}e\,\dd r+\frac{\Sigma J}{\mathcal F}\dd r^2
 \nonumber\\
 &+\frac{\Sigma}{b}\dd z^2+\frac{b}{\Xi^2\Sigma}f^2.
\end{align}
For $\mathcal F>0$ and smooth $u$, the metric coefficients are regular
at a simple horizon away from the polar-coordinate degeneracy.
$\mathcal D>0$ additionally makes the preferred leaves spacelike.
This chart permits reconstruction of $\gamma_{ij}$, $N$, $N^i$, and
$K_{ij}$ without differentiating a singular BL height.

On the axis, $b\simeq\Xi\theta^2$ and
$\gamma_{\theta\theta}$ and $\gamma_{\psi\psi}/\theta^2$ both approach
$(r^2+a^2)/\Xi$. Elementary flatness therefore holds for smooth $u$
and $2\pi$ periodicity of $\psi$; the factors $b$ and $1/b$ are
coordinate degeneracies, not extra physical boundaries.

Future orientation is fixed by the normal flux. At the future
black-hole horizon $n^r<0$ requires $P(r_+,z)>0$; at the future
cosmological horizon the outward normal requires $P(r_c,z)<0$.
For compatible reference data, $s(r_+)=+1$ and $s(r_c)=-1$, so
Eq.~\eqref{eq:regularmetric} interpolates between the corresponding
ingoing and outgoing regular charts. Reversing the signs selects the
past horizons, even if the coordinate coefficients stay finite.

\paragraph{Metric horizon and apparent-horizon terminology.}
The radii $r_+$ and $r_c$ are Killing horizons of the four-metric.
On a regular black-hole horizon section the outgoing null normal is
proportional to $\chi=\partial_t+[a/(r_+^2+a^2)]\partial_\psi$.
Its expansion vanishes because
$q_{\mathcal S}^{\mu\nu}\nabla_\mu\chi_\nu=0$ for this Killing field.
An apparent-horizon identification additionally requires an outermost
or trapped-region boundary characterization on the leaf
\cite{GourgoulhonJaramillo2006}. We do not perform that global search,
or infer the causal boundary of the constrained VCDM sector from the
metric horizon alone.

\section{Regularity of the VCDM clock and interior endpoints}\label{sec:zero}
\subsection{Localized lapse zeros and foliation endpoints}
\label{sec:clockregularity}
For the prescribed clock $T$, define
\begin{equation}\label{eq:clocknorm}
 X_T\equiv-g^{\mu\nu}\nabla_\mu T\nabla_\nu T=N^{-2}.
\end{equation}
If $N\to0$ at a point where the metric has a smooth nondegenerate
extension, $X_T\to\infty$ and the same $T$ cannot extend there as a
$C^1$ scalar. This statement is invariant under spacetime coordinate
changes. To classify the endpoint of the preferred foliation, one must
also distinguish time relabeling from regularity of the normalized normal.

For a differentiable relabeling $\widetilde T=f(T)$ with
$0<f'(T)<\infty$ on the open domain,
\begin{equation}\label{eq:clockrelabel}
 \widetilde N=\frac{N}{f'},\qquad
 X_{\widetilde T}=(f')^2X_T,\qquad
 \widetilde n_\mu=n_\mu.
\end{equation}
The normal and $K_{\mu\nu}$ are unchanged. A divergence of $X_T$ persists
under relabelings with finite positive $f'$ and $1/f'$ at the proposed
extension. In special cases a label degenerating uniformly on a limiting
leaf can nevertheless describe a regular normalized foliation.\footnote{
In Minkowski space with $\tau>0$, take $T=\sqrt{\ell\tau}$, where
$\ell>0$ is a constant length. Then $X_T=\ell/(4\tau)$ and
$N=2\sqrt{\tau/\ell}$, yet $n_\mu\dd x^\mu=-\dd\tau$,
$K_{ij}=0$, and $a_\mu=0$. The label $\widetilde T=T^2/\ell=\tau$
extends regularly. This lapse collapse is uniform on each leaf; the
example is outside the stationary ansatz $T=\tau+h(r,z)$ and does not
evade the BL acceleration obstruction.}
Such a relabeling need not preserve a prescribed boundary clock or give
an admissible extension of all VCDM fields.

\paragraph{The same-leaf obstruction.}
The crucial restriction is that $f'(T)$ is constant on each connected
leaf. For any two points $p,q$ on the same leaf,
\begin{equation}\label{eq:sameleafratio}
 \frac{\widetilde N(p)}{\widetilde N(q)}
 =\frac{N(p)}{N(q)},\qquad T(p)=T(q).
\end{equation}
Suppose $p$ approaches a lapse-zero end along a leaf while a reference
point $q$ on that leaf has finite positive lapse. No relabeling can make
the lapse finite and strictly positive at both the end and the reference
portion. The same argument applies to a limiting family of pairs with
$T(p)=T(q)$: even if $f'$ is allowed to degenerate as the leaf label approaches
a limiting value, compensating $N(p)\to0$ would spoil regularity at $q$
when $N(q)$ tends to a finite positive value. Spatial nonuniformity on the
same leaf, rather than whether a zero set moves in spatial coordinates,
is the relevant condition.

For our stationary normalization, let $\xi^\mu=(\partial_\tau)^\mu
=(\partial_t)^\mu$. Since $\xi^\mu\nabla_\mu T=1$,
\begin{equation}\label{eq:stationarylapseprojection}
 N=-n_\mu\xi^\mu.
\end{equation}
An inner zero approached along a stationary leaf, with positive lapse in
its exterior portion, is therefore not merely a leafwise clock-rescaling
artifact. This applies to a limiting graph $r=R(z)$ as well as to a
constant-radius limit. If the unit normal extends smoothly as the normal
to a regular limiting spacelike hypersurface, $N=-n_\mu\xi^\mu\to0$
implies that the stationary Killing field is tangent to that hypersurface.
This is a necessary tangency condition, not a proof that such an extension
exists or a classification of the endpoint's proper distance.

\paragraph{A finite-proper-distance criterion.}
The normal acceleration satisfies~\cite{Gourgoulhon2007}
\begin{equation}\label{eq:accelrelabel}
 a_i=D_i\ln N,\qquad
 \widetilde a_i=D_i\ln(N/f')=a_i,
\end{equation}
because $D_iT=0$. Let $\Gamma$ be a curve within a smooth spacelike
leaf, parametrized by proper length $\ell$, from a regular point $q$
to $p$. Its unit tangent and the spatial Cauchy--Schwarz inequality give
\begin{equation}\label{eq:finitepathlapse}
 \left|\ln\frac{N(p)}{N(q)}\right|
 \leq\int_{\Gamma(q,p)}\sqrt{a_i a^i}\,\dd\ell.
\end{equation}
If $N(p)\to0$ along such a curve of finite total proper length, the
integral must diverge: $a_i a^i$ cannot remain bounded. At a finite
regular spacetime endpoint this excludes a $C^1$ extension of the normal,
independently of time relabeling. A finite coordinate radius alone does
not supply the finite-length hypothesis. At an infinite-proper-distance
trumpet end, $N$ may instead tend to zero with bounded acceleration,
as the spherical example below shows.

We use \emph{endpoint} to describe the boundary or limiting end reached
by the stationary foliation, not to assert that it is regular or singular.
Nonremovability by $f(T)$ and local singularity are distinct statements.
The BL acceleration gives a genuine singularity of its normal; an
asymptotic trumpet need not be a local singularity at finite proper
distance. No rotating endpoint below $r_{\rm in}$ has been computed here.

\subsection{An exact exterior exclusion identity}
For the normalized stationary ansatz $T=\tau+h(r,z)$, the unit norm of
$n^\mu$ gives a useful kinematic identity, independently of the CMC
equation. Write
\begin{equation}\label{eq:W}
 \WW=\Xi^2\left[J^2-\frac{a^2(1-z^2)\Dr}{\Dz}\right]
      =\mathcal A\Dr.
\end{equation}
Multiplying
$\mathcal A N^2-\Dr N^2h_r^2-bN^2h_z^2=\Sigma$ by $\Dr$ and using
the flux definitions gives
\begin{equation}\label{eq:barrier}
 \boxed{\WW N^2=P^2+\Dr\left(\Sigma+\frac{\QQ^2}{b}\right).}
\end{equation}
On a smooth axis $\QQ=O(1-z^2)$, so $\QQ^2/b\to0$ regularly.
The rearrangement
\begin{equation}\label{eq:Wpositive}
 \WW=\frac{\Xi^2}{\Dz}
 \left[\Xi J\Sigma+a^2(1-z^2)(2Mr-q^2)\right]
\end{equation}
shows positivity when $r>0$, $\Sigma>0$, and $2Mr\ge q^2$, sufficient
conditions satisfied in the studied exteriors. Equivalently,
$g_{\phi\phi}=b\WW/(\Xi^4\Sigma)$, so $\WW>0$ means the axial
Killing vector is spacelike away from the axis. The AF inequality
$r_+>M\ge|q|$ for subextremal Kerr--Newman should not be assumed as
a general dS characterization; the sign can be checked directly.

In the regular $\Dr>0$ exterior block, Eq.~\eqref{eq:barrier} yields
\begin{equation}\label{eq:lowerbound}
 \boxed{N^2\ge\frac{\Sigma\Dr}{\WW}=N_{\rm BL}^2>0.}
\end{equation}
Thus the stationary preferred lapse cannot vanish at a finite regular
point in that block. The statement also holds in the ergoregion,
because the relevant sign is that of $\Dr$, not $g_{\tau\tau}$.
It excludes this particular endpoint without proving smoothness of all
normal derivatives. The numerical value of the bound uses
$\partial_\tau t=1$; a regular finite relabeling cannot create a lapse
zero where none exists.

With finite normal fluxes and $\WW>0$, a lapse zero can lie at a horizon
only if $P$ also vanishes, or in a $\Dr<0$ region where the two terms
on the right of Eq.~\eqref{eq:barrier} cancel. For an inward continuation
this points to the black-hole horizon or trapped block; the dS region
beyond the cosmological horizon must be checked separately. These are
necessary location conditions, not an assertion that an endpoint occurs.

\begin{figure}[tbp]
 \centering
 \includegraphics[width=\columnwidth]{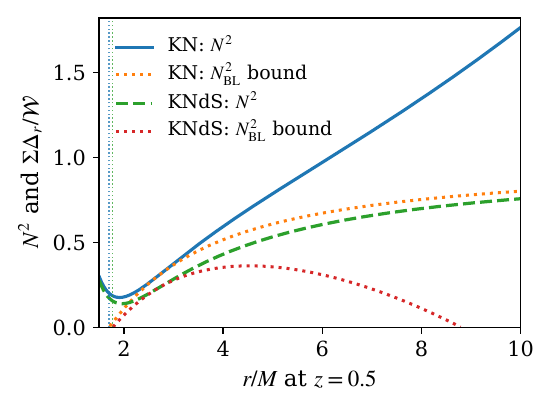}
 \caption{Squared lapse of the rotating benchmark $a=0.6$, $q=0.4$,
 $\kappa=0.1$, $C=2$ at $z=0.5$, and the analytic lower bound
 \eqref{eq:lowerbound}, in units $M=1$. The bound is shown only where
 $\Dr>0$: outside $r_+$ in AF and between $r_+$ and $r_c$ in dS.
 Numerical curves start at $r_{\rm in}=1.5$; dotted vertical lines mark
 the black-hole horizons. No interior lapse-zero extrapolation is drawn.}
 \label{fig:barrier}
\end{figure}

At a simple horizon, $\WW_h=\Xi^2J_h^2$ and the exact horizon lapse is
\begin{equation}\label{eq:horizonN}
 N(r_h,z)=\frac{|P(r_h,z)|}{\Xi(r_h^2+a^2)}.
\end{equation}
For a regular spacelike solution, strictly positive lapse and the
intended future orientation therefore require
\begin{equation}\label{eq:horizonconditions}
 \min_zP(r_+,z)>0,\qquad\max_zP(r_c,z)<0.
\end{equation}
The second condition applies only in dS. A nonzero absolute flux alone
does not select future rather than past horizons; a positive angular
mean alone does not enforce the pointwise sign.

Decompose $P(r_h,z)=C-\kappa r_hJ_h+\delta P_h(z)$, with
$\int_0^1\delta P_h\dd z=0$. Integrating the pointwise signs gives
the necessary conditions
\begin{equation}\label{eq:Cnecessary}
 C>\kappa r_+J_+,\qquad C<\kappa r_cJ_c\quad(\Le>0).
\end{equation}
A sufficient angular-fluctuation criterion is
\begin{align}\label{eq:Csufficient}
 C-\kappa r_+J_+&>\|\delta P_+\|_\infty,\\
 \kappa r_cJ_c-C&>\|\delta P_c\|_\infty.
\end{align}
For spherical leaves $\delta P_h=0$, necessity and sufficiency coincide.
For rotation, these inequalities protect the sign only when the angular
norm is controlled. The numerical tests below use sampled maxima,
not rigorous all-angle bounds.

\subsection{Where an inner limiting leaf can be located}
Suppose now that a solution satisfies the horizon-crossing conditions~\eqref{eq:horizonconditions} at $r_+$ and is continued inward into the black hole ($r < r_+$). If this continuation approaches a stationary limiting hypersurface away from a horizon ($\Dr \neq 0$) where the lapse vanishes ($N \to 0$) while the normal fluxes $(P, \QQ)$ remain finite and $P \neq 0$, setting $N = 0$ in the exact identity~\eqref{eq:barrier} yields
\begin{equation}\label{eq:zeroDelta}
 \Dr=-{P^2\over\Sigma+\QQ^2/b}<0.
\end{equation}
Because $\Dr < 0$ is required, any such inner limiting surface below $r_+$ must lie strictly inside the trapped black-hole region $r_- < r < r_+$ between the inner (Cauchy) and outer event horizons.

We can derive an even sharper geometric constraint if the stationary inner limiting leaf is described by a smooth axisymmetric graph $r = R(z)$ (the rotating generalization of a spherical trumpet cylinder $r = R_*$). A stationary surface $r - R(z) = 0$ (which is tangent to the Killing vectors $\partial_\tau$ and $\partial_\phi$) is spacelike if and only if its normal one-form $\nu_\mu = \partial_\mu[r - R(z)]$ is timelike, i.e., $g^{\mu\nu}\nu_\mu\nu_\nu < 0$. Using $g^{rr} = \Dr/\Sigma$, $g^{zz} = b/\Sigma$, and $g^{rz} = 0$, this condition reads
\begin{equation}\label{eq:graphbound}
 g^{\mu\nu}\partial_\mu(r-R)\partial_\nu(r-R)
 ={\Dr(R)+b[R'(z)]^2\over\Sigma(R,z)}<0.
\end{equation}
(Notice that the same inequality follows directly from Eq.~\eqref{eq:zeroDelta} when the level sets of $h(r, z)$ pile up along $r = R(z)$, where $\QQ/P = b h_z / (\Dr h_r) \to -b R'(z)/\Dr$.) Because $\Sigma(R, z) > 0$ and $b(z) = (1 - z^2)\Dz \ge 0$, Eq.~\eqref{eq:graphbound} requires both that $R(z)$ lie in the trapped block and that its polar slope $R'(z) = \dd R/\dd z$ satisfy the pointwise bound
\begin{equation}\label{eq:graphineq}
 r_-<R(z)<r_+,\qquad b[R'(z)]^2<-\Dr(R)
\end{equation}
throughout that block. Equations~\eqref{eq:zeroDelta}--\eqref{eq:graphineq} provide necessary geometric placement and slope bounds for any stationary spacelike inner limiting surface, without assuming that a given exterior solution actually reaches such a surface.

These bounds address an endpoint \emph{if it is reached}. They do
not establish that a lapse zero is the first singularity of every
rotating interior, or that all choices of $C$ and angular data have
such a zero. The solved domain stops at $r_{\rm in}$; the spherical
classification below already provides examples with no
positive-radius lapse zero. What is established here is regular
future-horizon crossing together with conditional placement of a
stationary spacelike limiting leaf. If a zero is approached along a
leaf with positive lapse in its exterior portion, the same-leaf
obstruction of Sec.~\ref{sec:clockregularity} applies. Its proper distance
and the limiting normal invariants must still be determined; the graph
bound alone does not classify the endpoint as singular or trumpet-like.

One important qualification is needed on the cosmological branch ($\Le > 0$). In Kerr--Newman--de Sitter spacetime, the region beyond the cosmological horizon ($r > r_c$) also has $\Dr < 0$. Therefore, the algebraic sign condition~\eqref{eq:zeroDelta} by itself does not automatically rule out a lapse zero in the far cosmological region $r > r_c$; rather, we check the cosmological side separately through the derived
$\rho\to0$ asymptotics, where $N\to\Xi^{-1}>0$, and positive
lapse values on numerical samples of the compactified exterior. These
checks are not an all-point regularity proof. Thus Eq.~\eqref{eq:graphineq} specifically characterizes an \emph{inward} limiting surface inside the black-hole block $r_- < r < r_+$.

\subsection{A complementary CMC positivity criterion}
Unlike the algebraic lapse identity, a second positivity criterion uses
stationarity of the CMC condition. Decompose
\begin{equation}\label{eq:Kdecomposition}
 K_{ij}=\sigma_{ij}+\frac{K_0}{3}\gamma_{ij},\qquad
 \gamma^{ij}\sigma_{ij}=0.
\end{equation}
CMC does not imply zero shear. For a hypersurface-orthogonal normal,
$a_i=D_i\ln N$ and $\nabla_\mu a^\mu=D^2N/N$. Raychaudhuri's equation
in our sign convention reads~\cite{Gourgoulhon2007}
\begin{equation}\label{eq:RaychaudhuriCMC}
 n^\mu\nabla_\mu K=-K_{ij}K^{ij}-R_{\mu\nu}n^\mu n^\nu
                         +\frac{D^2N}{N}.
\end{equation}
Because $K=K_0$ is constant in both space and time,
\begin{equation}\label{eq:lapseelliptic}
 D^2N=N(K_{ij}K^{ij}+R_{\mu\nu}n^\mu n^\nu).
\end{equation}
Equation~\eqref{eq:EM} gives
$R_{\mu\nu}n^\mu n^\nu=E_iE^i+B_iB^i-\Le$, with electric and
magnetic fields measured by $n^\mu$. Thus
\begin{align}\label{eq:lapseU}
 D^2N&=\mathcal U N,\\
 \mathcal U&=\sigmasq+3(\kappa^2-\HL^2)+E_iE^i+B_iB^i\ge0.
\end{align}
The sign holds for both computed branches: $\kappa=\HL$ in dS and
$\kappa>\HL=0$ in AF.

On a connected compact domain of a smooth Riemannian leaf with regular
coefficients and strictly positive boundary lapse, the strong maximum
principle implies $N>0$ in the interior. A noncompact end can be treated
by finite truncations where the known asymptotics give positive outer
boundary data. This excludes an isolated interior zero of a regular
stationary solution, but assumes the spatial geometry is already
regular. It neither proves existence for the nonlinear CMC problem nor
provides an extension through a degenerating leaf.

\subsection{Spherical reference: endpoints and proper depth}
\label{sec:sphericalzero}
An exact reference separates the location of a lapse-zero surface
from its intrinsic distance. For the spherically symmetric CMC foliation
subfamily~\cite{BrillCavalloIsenberg1980,IriondoMalecOMurchadha1996,MalecOMurchadha2003,MalecOMurchadha2009}
($a=0$), after fixing a constant time offset $u=0$, define
\begin{align}\label{eq:sphericals}
 f(r)&=1-\frac{2M}{r}+\frac{q^2}{r^2}-\frac{\Le r^2}{3},\\
 \beta(r)&=\frac{C}{r^2}-\kappa r,\qquad
 N^2=S(r)=f(r)+\beta(r)^2.\nonumber
\end{align}
The radius is areal, $A=4\pi r^2$, and the induced metric is
$\dd\ell^2=S^{-1}\dd r^2+r^2\dd\Omega^2$. With $n^r=-\beta$,
\begin{align}\label{eq:sphericalcomponents}
 K^\theta{}_{\theta}=K^\phi{}_{\phi}&=\kappa-\frac{C}{r^3},&
 K^r{}_r&=\kappa+\frac{2C}{r^3},
\end{align}
so that
\begin{equation}\label{eq:sphericalK}
 K_{ij}K^{ij}=3\kappa^2+\frac{6C^2}{r^6},\qquad
 \sigmasq=\frac{6C^2}{r^6}.
\end{equation}
These scalars can remain finite at $N=0$, although $X_T$ diverges.
Another, clock-relabeling-invariant diagnostic is the acceleration
of the preferred normal, $a_i=D_i\ln N$:
\begin{equation}\label{eq:accel}
 a_i a^i=\frac{[S'(r)]^2}{4S(r)}.
\end{equation}
The finite extrinsic-curvature invariant in the uncharged case was
already identified in Ref.~\cite{DeFeliceDollLarrouturouMukohyama2021}.

At a simple zero reached from $S>0$,
$S\simeq S'(R_0)(r-R_0)$ with $S'(R_0)>0$. The proper distance to the
endpoint is finite and the acceleration invariant diverges, realizing
the obstruction in Eq.~\eqref{eq:finitepathlapse}. If the background
metric is regular there, this is a singularity of the preferred normal,
not just of the time label. At a double root $R_*$ with $S''(R_*)>0$,
$S\simeq S''(R_*)(r-R_*)^2/2$, so instead
\begin{equation}\label{eq:trumpetaccellimit}
 a_i a^i\longrightarrow\frac{S''(R_*)}{2},\qquad
 \ell=\int\frac{\dd r}{\sqrt S}\longrightarrow\infty.
\end{equation}
The divergence of the distance is logarithmic. This is a trumpet:
a finite-area limiting surface at infinite proper spatial distance
\cite{BuchmanPfeifferBardeen2009,DennisonBaumgarte2017}. Both the simple and double zeros are
spatially localized and cannot be removed by $f(T)$ while keeping the
lapse finite and positive on the exterior portion of the same leaf.
Their geometric character nevertheless differs: the double root is an
asymptotic end with finite displayed normal invariants, not a demonstrated
local singularity at finite proper distance. Their finiteness does not,
by itself, prove extendibility of the complete VCDM configuration.

The location also depends on the foliation data. On the future
spherical branch, a zero in $f<0$ obeys
\begin{equation}\label{eq:Ccurve}
 C=\kappa R_0^3+R_0^2\sqrt{-f(R_0)}\equiv C_+(R_0).
\end{equation}
For $q=\kappa=\Le=0$, the critical value and radius are
\begin{equation}\label{eq:Schwarzschildcritical}
 C_*={3\sqrt3\over4}M^2,\qquad R_*={3M\over2}.
\end{equation}
For $0<C<C_*$ the outer positive root is simple; at $C=C_*$
it is double; for $C>C_*$ there is no positive-radius lapse zero.
Thus cosmological smallness does not itself select a trumpet or even
guarantee that $N$ vanishes before the central region is reached.

Following the future critical branch, with $q=0$ and both
$\kappa M$ and $|\Le|M^2$ small, the double-root equations of
Appendix~\ref{app:spherical} give
\begin{align}\label{eq:smallcosmotrumpet}
 {R_*\over M}&={3\over2}+{9\sqrt3\over8}\kappa M
 +O\!\left((\kappa M)^2,|\Le|M^2\right),\\
 {r_{\rm h}\over M}&=2+{8\over3}\Le M^2
 +O\!\left((\Le M^2)^2\right).\nonumber
\end{align}
Here $r_{\rm h}$ is the black-hole horizon of the frozen spherical
metric; $C_*$ is varied with the parameters to maintain the double
root. Hence $R_*/r_{\rm h}\simeq3/4$ and
$r_{\rm h}-R_*\simeq M/2$, despite an infinite proper distance on
the preferred leaf. In the Schwarzschild critical limit,
\begin{equation}\label{eq:criticaldistance}
 \ell(R_*+\delta,r_{\rm h})
 ={3M\over2\sqrt2}\ln\!\left({M\over\delta}\right)+O(M),
 \qquad\delta\to0^+.
\end{equation}
With $\kappa=\Le=0$ and $|q|<M$, the outer critical radius instead is
\begin{equation}\label{eq:RNcriticalradius}
 R_*={3M+\sqrt{9M^2-8q^2}\over4}.
\end{equation}
These are spherical critical references, not positions assigned to
the rotating numerical solutions or to a dynamically selected
cosmological endpoint.

\begin{figure}[tbp]
 \centering\includegraphics[width=\columnwidth]{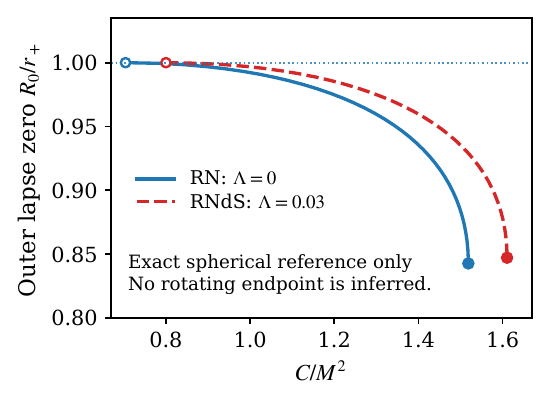}
 \caption{Exact spherical reference, $M=1$, $q=0.4$, $\kappa=0.1$.
 The outer lapse zero moves inward in areal radius as $C$ is
 increased. Open endpoints have zero horizon lapse and are excluded
 from regular crossing. Filled endpoints are double-root trumpets,
 at infinite proper spatial distance despite their finite areal
 depth. No rotating endpoint is inferred.}\label{fig:sphericalzero}
\end{figure}

For the finite-parameter reference $M=1$, $q=0.4$, $\kappa=0.1$,
the critical pairs are
\begin{align}\label{eq:sphericalcriticalvalues}
 (R_*,C_*)&\simeq(1.6150764,1.5186859),&&\Le=0,\\
 (R_*,C_*)&\simeq(1.6943901,1.6106981),&&\Le=0.03.
\end{align}
Figure~\ref{fig:sphericalzero} illustrates these spherical loci.
For a rotating inner graph, $r$ is not an areal radius, and a proper
distance must be defined using the induced $\gamma_{ij}$ and a
specified spatial path or distance-minimizing construction from the
horizon section. The bounds \eqref{eq:graphineq} constrain placement,
not that distance. An infalling observer's proper time is a different
quantity again.

\section{The two asymptotic boundary-value problems}\label{sec:asympt}
The outer conditions are part of the construction of the new
preferred-time configurations, not optional coordinate choices.
The cosmological expansion below makes the rotating charged CMC
family compatible with an expanding de Sitter end; the flat branch
requires a different hyperboloidal expansion. These derived data
supply the asymptotic part missing from a bare specification of the
Einstein--Maxwell metric.

To solve the stationary CMC equation~\eqref{eq:CMC} all the way to infinity without introducing an artificial outer boundary at finite radius, we compactify the exterior radial coordinate by setting
\[
 \rho=\frac{1}{r},
\]
so that $r \to \infty$ maps to the regular coordinate boundary $\rho = 0$. Although the bulk conservation law takes the same form on both VCDM branches, the asymptotic geometry of the slices as $\rho \to 0$ differs fundamentally between the expanding de Sitter branch ($\kappa = \HL > 0$) and the asymptotically flat hyperboloidal branch ($\HL = 0, \kappa > 0$). We therefore derive their asymptotic expansions and $\rho = 0$ boundary conditions separately.

\subsection{Expanding de Sitter end}\label{sec:dSclock}
On the cosmological branch we have $\kappa = \HL > 0$, so the mean curvature $K_0 = 3\kappa = 3\HL$ matches the asymptotic expansion rate of spatially flat slices in de Sitter space.

It is helpful to examine first the pure de Sitter reference limit
obtained by setting $M = q = C = 0$ while keeping the Boyer--Lindquist
oblate parameter $a \neq 0$. In this limit, $\bar P = -\kappa r J$ and
$\Dr = J(1 - \kappa^2 r^2)$, so the reference function in
Eq.~\eqref{eq:radialreference} simplifies to
$\mathcal F = \bar P^2 + \Dr J = J^2$ and $s = -\kappa r$. Because
$(r, z)$ with $a \neq 0$ are oblate spheroidal coordinates rather than
standard spherical coordinates, a spatially flat slicing of pure de
Sitter space requires a nontrivial angular dependence in the residual
height $u$. Specifically, one may verify directly from
Eqs.~\eqref{eq:D} and \eqref{eq:CMC} that the exact pure de Sitter
solution in our regularized variables is
\begin{equation}
 u_{\rm dS}=\frac{\Xi}{2\kappa}\log\Dz,\qquad
 N_{\rm dS}=\Xi^{-1},
\end{equation}
The constant reference lapse also fixes the relation between our
coordinate time and proper cosmological time. Along the normal
congruence of this pure de Sitter reference,
\begin{equation}\label{eq:cosmicclock}
 t_{\rm cos}=\frac{t}{\Xi},\qquad
 N_{\rm cos}=\Xi N,\qquad N_{{\rm cos},\infty}=1,
\end{equation}
up to an additive time origin. This constant relabeling changes
neither the leaves nor $n^\mu$ and $K_0=3H_\Lambda$.
With a black hole present it specifies the asymptotic time
normalization; it does not make $t_{\rm cos}$ proper time at every
spatial point. We retain the original $t$ normalization in the
solver and all numerical tables. On the flat branch,
$N\sim\kappa r$ instead, so no constant rescaling makes the
hyperboloidal coordinate a globally comoving FLRW proper time.

The corresponding normal fluxes are
\[
 P_{\rm dS}=-\kappa rJ,\qquad \QQ_{\rm dS}=\kappa a^2z(1-z^2).
\]
Notice that $\partial_r P_{\rm dS} + \partial_z \QQ_{\rm dS} = -\kappa(3r^2 + a^2) + \kappa a^2(1 - 3z^2) = -3\kappa\Sigma$, so the CMC equation~\eqref{eq:CMC} is satisfied identically. We emphasize that the nonvanishing polar flux $\QQ_{\rm dS}$ is purely a geometric consequence of using oblate coordinates $(r, z)$ with $a \neq 0$, not a physical rotating source in pure de Sitter spacetime.

When the black-hole mass $M$, charge $q$, and flux constant $C$ are turned on, the leading cancellation in $\mathcal F = \bar P^2 + \Dr J = J^2 - 2(M + \kappa C)r J + (q^2 J + C^2)$ is governed by the combined mass--flux parameter (of dimension length)
\[
 B\equiv M+\kappa C.
\]
Searching for a smooth power-series solution around $\rho = 0$, we expand $u(\rho, z)$ beyond the pure de Sitter reference as
\begin{equation}\label{eq:dsasym}
 u=u_{\rm dS}+\rho u_1(z)+\rho^2u_2(z)+O(\rho^3).
\end{equation}
To see how $u_1(z)$ is fixed, note that substituting Eq.~\eqref{eq:dsasym} into the lapse~\eqref{eq:D} gives $N = \Xi^{-1}[1 - (B - \kappa u_1/\Xi)\rho + O(\rho^2)]$. In turn, the polar flux $\QQ = N b u_z$ expands as
\begin{align*}
 \QQ&=\QQ_{\rm dS}+\rho\,Q_1(z)+O(\rho^2),\\
 Q_1(z)&=\frac{b}{\Xi}u_1'(z)
 -\kappa a^2z(1-z^2)\left[B-\frac{\kappa u_1(z)}{\Xi}\right].
\end{align*}
Define $\mu(z)=B-\kappa u_1(z)/\Xi$. On the branch where the
subtracted radial flux has a finite power-series limit, the leading
compactified conservation law and regularity at the equator and axis
set the coefficient of $\rho$ in $\QQ-\QQ_{\rm dS}$ to zero.
The displayed coefficient then gives
\begin{equation}
 \mu'(z)=-\frac{\kappa^2a^2z}{\Dz}\mu(z),
 \qquad \mu(z)=\frac{\mu_0}{\sqrt{\Dz}}.
\end{equation}
The integration constant is fixed by the mean flux, not by setting
$u_1$ to zero on the axis. Direct expansion of the radial flux yields
\begin{equation}
 \Pi(0,z)=-\frac{M}{\kappa}
       +\frac{\Xi\mu_0}{\kappa\Dz^{3/2}},
 \qquad \int_0^1\Pi(0,z)\dd z=C.
\end{equation}
Since $\int_0^1\Dz^{-3/2}\dd z=\Xi^{-1/2}$, it follows that
$\mu_0=(M+\kappa C)/\sqrt\Xi=B/\sqrt\Xi$. Therefore
\begin{equation}\label{eq:u1}
 u_1=\frac{\Xi B}{\kappa}\left[1-(\Xi\Dz)^{-1/2}\right],
\end{equation}
which yields the asymptotic lapse expansion
\begin{equation}\label{eq:Nasym}
 N=\frac1\Xi\left[1-\frac{B\rho}{\sqrt{\Xi\Dz}}+O(\rho^2)\right].
\end{equation}
In particular, $u_1(1)=\kappa a^2B$, which is generally nonzero;
axis smoothness does not prescribe this value. Proceeding to the next
order in $\rho$ determines the quadratic coefficient $u_2(z)$ (which incorporates the leading effect of the electric charge $q^2$); its closed-form expression and the second-order angular differential operator that governs it are presented in Appendix~\ref{app:asymptotics}.

This asymptotic structure motivates defining the pure-de Sitter-subtracted normal fluxes
\begin{equation}\label{eq:regflux}
 \Pi=P+\kappa rJ,\qquad
 \Theta=\frac{\QQ-\kappa a^2z(1-z^2)}{\rho^2}.
\end{equation}
Because $P_{\rm dS} = -\kappa r J$ and $\QQ_{\rm dS} = \kappa a^2 z(1 - z^2)$ already balance the source term $-3\kappa\Sigma$ in Eq.~\eqref{eq:CMC}, rewriting $\partial_r = -\rho^2\partial_\rho$ transforms the bulk CMC equation~\eqref{eq:CMC} into the exact, source-free compactified conservation law
\begin{equation}\label{eq:compact}
 \mathcal R_{\rm comp}\equiv-\partial_\rho\Pi+\partial_z\Theta=0.
\end{equation}
On the smooth asymptotic branch~\eqref{eq:dsasym}, both $\Pi(\rho, z)$ and $\Theta(\rho, z)$ approach finite, smooth limits as $\rho \to 0$. In our numerical solver, we factor out the known leading asymptotics by solving for the rescaled exterior unknown
\[
 v(\rho,z)\equiv\frac{u(\rho,z)-u_{\rm dS}(z)-\rho u_1(z)}{\rho^2},
\]
subject to the exact Dirichlet boundary condition $v(0, z) = u_2(z)$ at $\rho = 0$. Notice that if one instead truncated the domain at a large finite radius $r_{\rm max}$ and naively imposed $u(r_{\rm max}, z) = u_{\rm dS}(z)$, one would omit the anisotropic $O(1/r)$ correction $\rho u_1(z)$ and inject a systematic $O(1/r_{\rm max})$ angular boundary error into the solution.

By working in the compactified coordinate $\rho \in [0, 1/R_m]$, the asymptotic boundary $\rho = 0$ is enforced directly on the computational grid. At the same time, we note a standard geometric caveat: the formal asymptotic expansion and numerical control of the
compactified endpoint do not establish a globally smooth Penrose
completion. The future conformal boundary of de Sitter spacetime is
spacelike, unlike the future null infinity of the asymptotically flat
branch.

\subsection{Asymptotically flat end}
On the asymptotically flat branch we set $\HL = 0$ (so $\Xi = \Dz = 1$ and $b = 1 - z^2$), while keeping $\kappa = K_0/3 > 0$. Physically, the four-dimensional Kerr--Newman spacetime is asymptotically flat, but because the slices have strictly positive constant mean curvature $K_0 = 3\kappa > 0$, they do not approach standard Euclidean spatial infinity ($i^0$, where $K \to 0$). Instead, the constant-$t$ hypersurfaces are \emph{hyperboloidal} slices that bend upward to intersect future null infinity ($\IR$)~\cite{Zenginoglu2008}.

To understand the asymptotic expansion of $u(\rho, z)$ in this regime, observe that when $\HL = 0$ and $\kappa > 0$, the term $\bar P^2 \sim \kappa^2 r^6$ dominates over $\Dr J \sim r^4$ in Eq.~\eqref{eq:radialreference}, so $\mathcal F = \bar P^2 + \Dr J \sim \kappa^2 r^6$ (rather than $r^4$ as in de Sitter). Consequently, $J^3/\mathcal F \to \kappa^{-2}$ and $\mathcal D \to \kappa^{-2}$ remain $O(1)$ at large $r$, while $\Sigma \sim r^2$. From $N = \sqrt{\Sigma/\mathcal D}$, the physical lapse grows linearly with radius, $N \sim \kappa r$,\footnote{As a simple example, consider
Minkowski spacetime~\cite{Zenginoglu2008}
$ds^2=-d\tau^2+dr^2+r^2d\Omega_2^2$ and introduce the
hyperboloidal time coordinate
$t=\tau-\sqrt{r^2+\kappa^{-2}}$, with $\kappa>0$.
The constant-$t$ hypersurfaces have $K=3\kappa$,
in our sign convention, and their lapse satisfies
$N^2=1+\kappa^2r^2$.
Thus $N\sim\kappa r$ as $r\to\infty$, even though
the underlying spacetime is exactly flat.
The conformal lapse $\bar N=N/r$, by contrast,
approaches the finite value $\kappa$.} which means that the conformally rescaled lapse
\[
 \bar N\equiv \rho N \to \kappa > 0 \qquad (\rho\to0)
\]
is finite and positive at $\rho = 0$. Likewise, in the compactified four-metric $\rho^2 \dd s^2$, the boundary $\rho = 0$ has a null normal, matching the standard conformal geometry of $\IR$.

What about the residual height function $u(\rho, z)$? Even in flat Minkowski spacetime ($M = q = C = 0$), our radial reference height $h_{\rm ref}'(r) = Js/\Dr = -\kappa r/\sqrt{1 + \kappa^2 r^2}$ integrates to $h_{\rm ref}(r) = -\sqrt{\kappa^{-2} + r^2}$ in terms of the Boyer--Lindquist oblate radius $r$, whereas a true spherically symmetric CMC hyperboloid in flat space is given by $t - \tau = -\sqrt{\kappa^{-2} + R_{\rm sph}^2}$ in terms of the spherical radius $R_{\rm sph} = \sqrt{r^2 + a^2(1 - z^2)}$. Expanding the difference $-\sqrt{\kappa^{-2} + r^2 + a^2(1 - z^2)} + \sqrt{\kappa^{-2} + r^2}$ in powers of $\rho = 1/r$ shows that the leading $O(\rho)$ and $O(\rho^3)$ terms are purely coordinate-conversion terms that depend only on $a$ and $\kappa$, while the $O(\rho^2)$ term vanishes identically. When $M$, $q$, and $C$ are nonzero, their effects enter starting at order $\rho^4$. Thus, on the smooth hyperboloidal branch, $u(\rho, z)$ has the asymptotic expansion
\begin{equation}\label{eq:afasym}
 u=\rho u_1(z)+\rho^3u_3(z)+\rho^4w(\rho,z),
\end{equation}
where the universal coordinate-geometry coefficients are
\begin{align}
 u_1&=-\tfrac12a^2(1-z^2),\\
 u_3&=\frac{a^2(1-z^2)}{4\kappa^2}
       +\frac{a^4(1-z^2)^2}{8},
\end{align}
and $w(\rho, z)$ is the smooth exterior unknown to be solved for numerically.

Just as in the de Sitter case, introducing the subtracted fluxes $\Pi$ and $\Theta$ via Eq.~\eqref{eq:regflux} transforms the CMC equation into the exact compactified conservation law~\eqref{eq:compact}, $-\partial_\rho\Pi + \partial_z\Theta = 0$, and both $\Pi$ and $\Theta$ remain finite as $\rho \to 0$. However, the boundary condition at $\rho = 0$ for the exterior unknown $w(\rho, z)$ is qualitatively different from the Dirichlet condition in de Sitter. At $\rho = 0$, the radial first integral~\eqref{eq:firstint} fixes only the angular average $\int_0^1 \Pi(0, z)\,\dd z = C$, not the pointwise angular profile $\Pi(0, z)$. In a rotating spacetime, the asymptotic flux $\Pi(0, z)$ at $\IR$ acquires a genuine angular dependence determined by the global solution. Therefore, rather than artificially forcing $w(0, z)$ to make $\Pi(0, z) = C$ isotropic pointwise, we enforce the regular limiting compactified PDE~\eqref{eq:compact} itself along the boundary row $\rho = 0$ with the flux mean inherited from the inner boundary and the
height offset fixed by the stated asymptotic convention. As we verify in our numerical tests, imposing an artificial isotropic flux $\Pi(0, z) = C$ at $\rho = 0$ produces deceptive collocation-node convergence while generating large off-grid PDE residuals near infinity; retaining the limiting PDE at $\rho = 0$ is thus essential for spectral accuracy.

\section{Numerical construction and error control}\label{sec:method}
For the original survey and the fixed-$\kappa$ flux scan we use a
two-domain pseudospectral collocation method. The small-$\kappa$ extension
uses the same physical equations and boundary conditions with the
additional numerical treatments described below and in
Appendix~\ref{app:newscans}. In the two-domain formulation the radial
coordinate is partitioned at $r=R_m$ into:
\begin{enumerate}
 \item a \emph{finite inner domain} $r_{\rm in} \le r \le R_m$, which starts at an excision radius $r_{\rm in}$ just inside the future outer black-hole horizon $r_+$ ($r_- < r_{\rm in} < r_+$) and extends into the exterior; and
 \item a \emph{compactified outer domain} $0 \le \rho \le 1/R_m$ in the inverse radius $\rho = 1/r$, which reaches asymptotic infinity at $\rho = 0$.
\end{enumerate}
In the inner domain we use a logarithmically mapped Chebyshev radial grid to concentrate resolution near the black-hole horizon, while in the outer domain we use a Chebyshev grid directly in $\rho$. In the polar direction $z = \cos\theta \in [0, 1]$, both domains expand the solution in an even-parity Chebyshev basis in $z$, which automatically enforces equatorial reflection symmetry ($u_z(r, 0) = 0$) and smoothness at the rotation axis ($z = 1$). Because the CMC equation~\eqref{eq:CMC} is a second-order elliptic equation in divergence form, the natural transmission conditions across the interface $r = R_m$ ($\rho = 1/R_m$) are pointwise continuity of the residual height $u$ and pointwise continuity of the radial normal flux $P$. The discretized inner and outer nonlinear algebraic systems, coupled through these interface conditions, are solved simultaneously as a single global system.

\subsection{Inner flux data and the family being computed}\label{sec:innerdata}
At the inner cutoff $r=r_{\rm in}$ we use the pointwise prescription
\begin{equation}\label{eq:innerbc}
 P(r_{\rm in},z)=C-\kappa r_{\rm in}(r_{\rm in}^2+a^2)
               =\bar P(r_{\rm in}).
\end{equation}
This is an additional boundary condition, not a consequence of the
CMC equation or of the VCDM auxiliary equations. The first integral
\eqref{eq:firstint} determines only the angular average. Setting the
zero-mean part $\delta P=P-\bar P$ to zero on one inner surface is a
simple, reproducible choice that introduces no extra angular
parameters and connects directly to the spherically symmetric flux
prescription. No interior regularity or collapse result derived here
selects this choice.

The quantity made angle-independent is the \emph{densitized coordinate
flux} $P$, not the normal velocity $n^r=-P/\Sigma$ or a flux per unit
proper area. For $a\ne0$, $\Sigma$ varies with angle even when $P$
does not. Nor does \eqref{eq:innerbc} impose $\QQ=0$, $u_z=0$, or
spherical symmetry of the slice. In fact, the bulk equation gives
\begin{equation}\label{eq:deltapevolution}
 \partial_r\delta P
 =\partial_z\left[\kappa a^2z(1-z^2)-\QQ\right].
\end{equation}
Thus angular structure generally develops away from $r_{\rm in}$,
although its mean remains zero at every radius.

More general data with the same value of $C$ have the form
\begin{align}\label{eq:generalinnerbc}
 P(r_{\rm in},z)&=\bar P(r_{\rm in})+g(z),&
 \int_0^1g(z)\dd z&=0,\\
 g(-z)&=g(z),&&
\end{align}
where $g$ is smooth at the axis and has dimensions of length squared.
For example,
\begin{equation}\label{eq:quadrupolebc}
 g(z)=\epsilon\,\mathsf P_2(z)
     =\frac{\epsilon}{2}(3z^2-1)
\end{equation}
preserves the mean flux, equatorial symmetry, and axis regularity.
Even Legendre multipoles $\mathsf P_{2\ell}$ with $\ell\ge1$ provide
further possible zero-mean profiles. Such data must still yield an
admissible spacelike solution with the required horizon signs and
asymptotic behavior: arbitrary profiles are not guaranteed to do so.
The freedom to choose angular inner data is also familiar from CMC
slicing in GR, including nonspherical CMC slices of a spherical
background~\cite{SchinkelMacedoAnsorg2014}.

For clarity, ellipticity does not disappear merely because this
boundary lies behind the metric horizon. Linearizing the stationary
fluxes under $h\mapsto h+\eta$ gives
\begin{align}\label{eq:fluxlinearization}
 \begin{pmatrix}\delta P\\\delta\QQ\end{pmatrix}
 &=\mathsf A\begin{pmatrix}\eta_r\\\eta_z\end{pmatrix},\nonumber\\
 \mathsf A&=N\begin{pmatrix}
 \Dr+P^2/\Sigma&P\QQ/\Sigma\\
 P\QQ/\Sigma&b+\QQ^2/\Sigma
 \end{pmatrix}.
\end{align}
Using \eqref{eq:barrier}, its determinant is
$\det\mathsf A=N^4b\WW/\Sigma$. Away from the coordinate axis,
$N>0$, $b>0$, and $\WW>0$ make this symmetric matrix positive
definite. A flux condition at the inner boundary is therefore part
of the stationary elliptic problem, not a hyperbolic excision rule
that can be omitted by placing a surface inside $r_+$.

Changing $g(z)$, or moving $r_{\rm in}$ while resetting $g=0$, generally
changes the boundary-value problem; the change need not be small.
To represent the same solution at a different cutoff, one would need
to impose its actual flux profile on the new surface. Likewise,
fixing $C$ and the leading asymptotic state does not fix every
subleading angular datum of the resulting solution. Under an
invertible linearized boundary-value operator, small compatible data
changes would admit a local continuation, but we do not prove that
hypothesis for the full compactified problem. The explicit test in
Sec.~\ref{sec:boundarytest} supplies limited numerical evidence instead.
Different solutions obtained for different $g$ are not a proof of
nonuniqueness for \emph{identical complete} boundary data, nor an
infinite set of independently observable black-hole hairs. The
extension below $r_{\rm in}$ and dynamical selection remain separate
questions.

\subsection{Nonlinear solution and numerical diagnostics}

We solve the coupled nonlinear collocation equations using Newton--Raphson iteration with analytical/differentiated flux Jacobians. At each Newton step, a backtracking line search monitors the lapse denominator $\mathcal D$ in Eq.~\eqref{eq:D} and rejects any trial update that would violate the spacelike slicing condition $\mathcal D > 0$. In the compactified outer domain $\rho \in [0, 1/R_m]$, evaluating the subtracted fluxes $\Pi$ and $\Theta$ in Eq.~\eqref{eq:regflux} naively from the unsubtracted fluxes $(P, \QQ)$ would suffer from severe floating-point cancellation near $\rho = 0$, where individual terms diverge like $r^3 = \rho^{-3}$. To avoid this numerical loss of significance, our exterior residual evaluators subtract the known leading Taylor coefficients analytically before dividing by powers of $\rho$. Of course, using a finite-order asymptotic Taylor jet near $\rho = 0$ is itself a controlled numerical approximation rather than an exact all-order evaluation; we therefore cross-checked the jet evaluators against higher-order expansions and direct arbitrary-precision (`mpmath`) evaluations of the unexpanded flux formulas.

Collocation residuals measure the discrete equations only. We also
sample continuous PDE residuals in both domains and use the following
complementary diagnostics:
\begin{itemize}
 \item verification of the radial first integral~\eqref{eq:firstint} across the domain and continuity of $u$ and $P$ at the matching interface $r = R_m$;
 \item verification of the horizon flux orientations~\eqref{eq:horizonconditions} and strict positivity of the lapse denominator $\mathcal D > 0$;
 \item an independent four-dimensional geometric reconstruction that builds the regular metric~\eqref{eq:regularmetric}, computes its Christoffel symbols by direct differentiation, and evaluates the extrinsic curvature tensor $K_{ij}$ and its trace $K = \gamma^{ij}K_{ij}$ without using the flux form of the PDE;
 \item point evaluations in extended precision verifying that the background metric and Maxwell field satisfy the four-dimensional Einstein--Maxwell--$\Le$ equations~\eqref{eq:EM}; and
 \item analytical benchmark tests in the exact spherically symmetric ($a = 0$), pure de Sitter ($M = q = C = 0, \kappa = \HL$), and oblate Minkowski ($M = q = C = 0, \HL = 0$) limits.
\end{itemize}

When interpreting the outer-domain accuracy near $\rho = 0$, it is important to distinguish the physical mean-curvature error $K - K_0$ from the regularized compactified residual $\mathcal R_{\rm comp} = -\partial_\rho\Pi + \partial_z\Theta$ defined in Eq.~\eqref{eq:compact}. Substituting $P = \Pi - \kappa r J$ and $\QQ = \rho^2\Theta + \kappa a^2 z(1 - z^2)$ into $K = -\Sigma^{-1}(\partial_r P + \partial_z\QQ)$ and using $\Sigma = \rho^{-2}(1 + a^2 z^2\rho^2)$ yields the exact relation
\begin{equation}\label{eq:errors}
 K-K_0=-\frac{\rho^4}{1+a^2z^2\rho^2}\mathcal R_{\rm comp}.
\end{equation}
Because of the prefactor $\rho^4 = r^{-4}$, a very small physical error $|K - K_0|$ near infinity does not automatically guarantee a small regularized residual $|\mathcal R_{\rm comp}|$; conversely, bounding $|\mathcal R_{\rm comp}|$ up to $\rho = 0$ is a much more stringent test that controls the higher-order asymptotic derivatives of the foliation. In our de Sitter benchmark ($a = 0.6, q = 0.4, \kappa = \HL = 0.1, C = 2$), an independent multiprecision audit of the double-precision spectral interpolant revealed a nonmonotonic residual plateau in $\mathcal R_{\rm comp}$ near $\rho = 0$, which was reduced to approximately $5 \times 10^{-8}$ on a fixed off-grid probe set after performing nonlinear Newton updates in extended precision. The broader parameter survey reported in Sec.~\ref{sec:results} uses standard double-precision runs with their own (generally larger) sampled residual maxima, and we do not claim uniform or monotonic spectral convergence of $\mathcal R_{\rm comp}$ as $\rho \to 0$. On the asymptotically flat hyperboloidal branch, the benchmark solution reaches off-grid regularized residuals $|\mathcal R_{\rm comp}|$ of order $10^{-10}$.

\paragraph{Continuous residuals at small cosmological curvature.}
For $\kappa M=0.01$, small collocation residuals alone proved insufficient.
If $\mathcal I_N$ denotes interpolation on the solution grid, the derivative
of $\mathcal I_N\Pi$ need not approximate the derivative of the nonlinear
flux $\Pi[\mathcal I_N u]$ accurately enough. We therefore distinguish the
algebraic residual solved by Newton from the continuous flux derivatives
of the represented state, including one-sided limits on interface rows
replaced by transmission conditions. The local physical mean-curvature
error and the compactified residual remain separate diagnostics as in
Eq.~\eqref{eq:errors}.

The final AF small-$\kappa$ calculation evaluates finite-$\rho$ fluxes
directly with multiprecision arithmetic and retains the analytic regular
PDE limit at $\rho=0$. It does not replace the physical AF boundary
condition by an isotropic flux condition. The final dS calculation splits
the compactified exterior into two subdomains, oversamples the nonlinear
fluxes before differentiation, and minimizes the de-aliased PDE residual
subject to the original boundary and transmission equations as numerical
equality constraints. The far-side interface PDE, whose collocation row
would otherwise be replaced, is included as an additional residual, not
as a new physical boundary condition. Acceptance is decided by the
original residual thresholds and separate interface and grid checks,
not by the least-squares objective alone. Appendix~\ref{app:newscans}
specifies the final configurations, arithmetic, and the scope of these
tests; failed intermediate iterates are not included as results.

Finally, we checked sensitivity to the choice of matching radius $R_m$. Unlike the inner cutoff $r_{\rm in}$ (which is a physical boundary of the computational domain where Eq.~\eqref{eq:innerbc} is imposed), the sphere $r = R_m$ is a purely numerical interface between the two spectral grids, so the underlying smooth solution is independent of $R_m$. In the asymptotically flat benchmark, solutions computed with $R_m = 32M$ and $R_m = 48M$ agree closely across all diagnostics. An early trial at $R_m=64M$ with the original implementation was
ill-conditioned and was not used in the survey comparisons. This is not
an existence restriction on the boundary-value problem; the supplementary
small-$\kappa$ calculations use different, larger matching radii with
controlled residual evaluation. The explicit numerical resolution parameters and acceptance thresholds for all seventeen survey configurations are summarized in Appendix~\ref{app:survey}.

\section{New rotating VCDM configurations and horizon regularity}\label{sec:results}
\subsection{The nonmaximal rotating charged families}
Using the two-domain spectral method described in Sec.~\ref{sec:method}, we have constructed and validated a survey of seventeen stationary, axisymmetric CMC black-hole configurations in VCDM: nine on the asymptotically flat hyperboloidal branch ($\Le = 0, \kappa > 0$) and eight on the expanding cosmological de Sitter branch ($\Le = 3\kappa^2 > 0$). Across this survey (summarized in Table~\ref{tab:S8}), the dimensionless spin parameter reaches $a/M = 0.8$, the electric charge reaches $q/M = 0.6$, and dedicated runs explore variations of the mean-curvature scale $\kappa M$ and the normal-flux constant $C/M^2$. In every configuration, the inner excision sphere $r_{\rm in}$ (chosen at $1.5M$ for moderate spin and charge, and shifted inward to $1.3M$ for the $a/M = 0.8$ and $q/M = 0.6$ cases) lies strictly inside the trapped region between the inner and outer black-hole horizons, $r_- < r_{\rm in} < r_+$, while the outer domain extends to compactified infinity $\rho = 0$. All seventeen configurations satisfy our sampled off-grid acceptance criteria, and their frozen states have been independently re-evaluated in audit reruns; this does not imply seventeen fresh nonlinear solves in the second implementation.

A practical remark regarding our choice of parameters is in order. In the principal cosmological configurations, we set $\kappa M = \HL M = 0.1$ (corresponding to $\Le M^2 = 0.03$). This value is chosen to provide a clean, numerically well-resolved testbed in which the black-hole and cosmological horizons are separated by less than an order of magnitude in radius, rather than to model the minuscule cosmological constant of our late-time universe around an astrophysical stellar- or supermassive-scale black hole (where $\HL M \ll 1$). The supplementary scan extends the comparison to $\kappa M=0.01$
using the methods described in Appendix~\ref{app:newscans}. Neither this
finite range nor the original survey resolves astrophysical scale
separations or the near-extremal limit $r_-\to r_+$. The seventeen selected configurations pass the stated sampled tests; the survey does not map the boundary of existence in parameter space.

To illustrate the basic geometry of our solutions, consider the primary rotating charged benchmark with parameters $M = 1$, $a = 0.6$, $q = 0.4$, $\kappa = 0.1$, and $C = 2$:
\begin{itemize}
 \item On the \emph{asymptotically flat Kerr--Newman branch} ($\Le = 0$), the outer black-hole horizon $\Dr(r_+) = 0$ is located at $r_+ = (1 + \sqrt{0.48})M \simeq 1.692820323M$.
 \item On the \emph{cosmological Kerr--Newman--de Sitter branch} ($\Le = 0.03$), the cosmological repulsion shifts the outer black-hole horizon slightly outward to $r_+ \simeq 1.767336796M$ and introduces a future cosmological horizon at $r_c \simeq 8.809499068M$.
\end{itemize}
Evaluating the preferred lapse function at a representative exterior test point $(r, z) = (3M, 0.5)$, we find $N \simeq 0.613796$ on the asymptotically flat branch and $N \simeq 0.534384$ on the cosmological branch. Notice, however, that passing from the first benchmark to the second changes \emph{both} the background four-dimensional spacetime metric (via $\Le$) and the preferred foliation; to isolate the effect of the foliation alone at fixed four-geometry, one must instead vary the flux parameter $C$ at fixed $(M, a, q, \Le, K_0)$, as we do in Sec.~\ref{sec:shear_results} below.

\subsection{Two-dimensional structure of the preferred foliation}
\label{sec:foliation2d}
\begin{figure*}[t]
 \centering
 \includegraphics[width=0.98\textwidth]{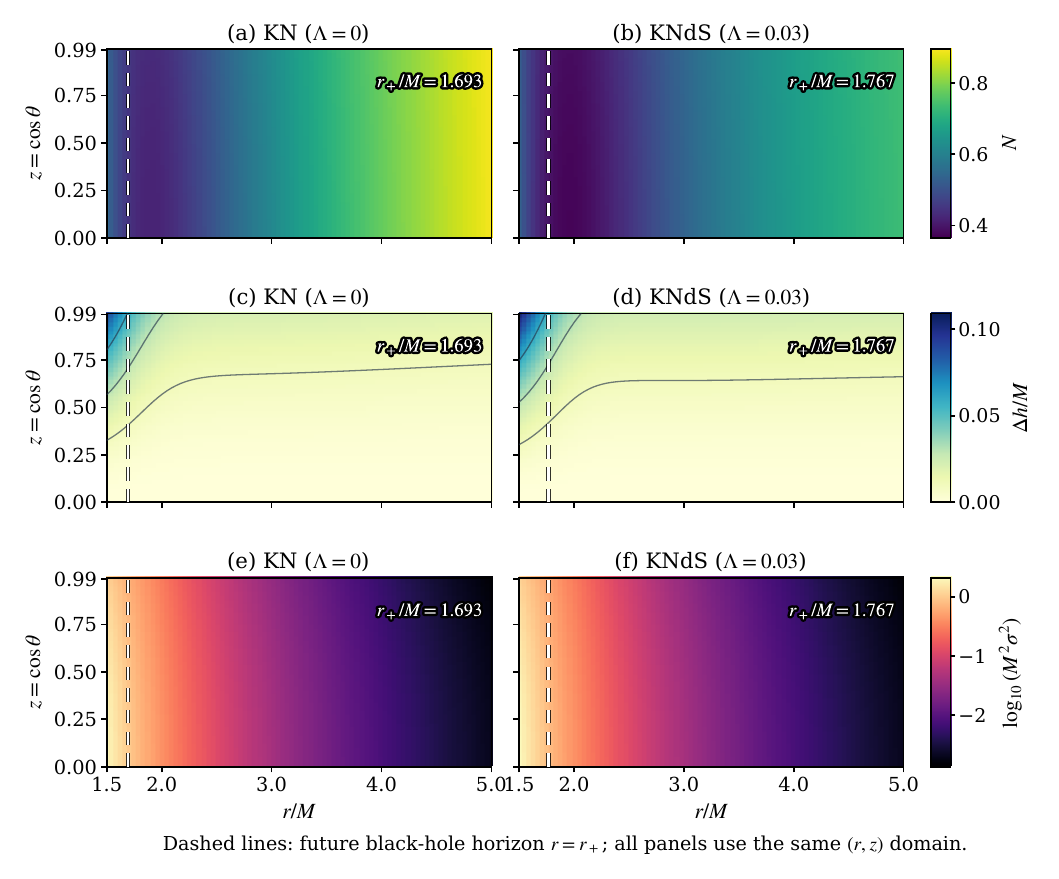}
 \caption{Meridional maps of the preferred foliation for the two rotating,
 charged benchmarks with $M=1$, $a=0.6$, $q=0.4$, $\kappa=0.1$, and $C=2$.
 Columns: Kerr--Newman ($\Le=0$) and Kerr--Newman--de Sitter
 ($\Le=0.03$). Rows: the lapse $N$ in the numerical time
 normalization; the angular height difference
 $\Delta h=h(r,z)-h(r,0)$ in the chosen reference chart; and the
 preferred-foliation scalar $\log_{10}(M^2\sigmasq)$.
 Each pair uses a common color scale; solid contours in the middle
 row mark $\Delta h/M=0.01,0.03,0.06$. Dashed lines mark the future
 black-hole horizons $r_+$, not lapse zeros. The maps cover
 $1.5\le r/M\le5$ and $0\le z\le0.995$; the polar-coordinate
 degeneracy excludes the exact axis $z=1$ from the Christoffel
 reconstruction. The cosmological horizon $r_c\simeq8.809M$ lies
 outside this near-hole view. No extrapolation below the
 excision surface or isometric embedding is implied.}
 \label{fig:foliation2d}
\end{figure*}

The geometry of the preferred slicing is not exhausted by a radial
lapse profile or a single-point shear diagnostic.  Figure~\ref{fig:foliation2d}
shows meridional maps reconstructed from the two principal rotating,
charged solutions, using the same radial and angular coordinates
$(r,z)$ on their respective Kerr--Newman backgrounds.  The first row
shows the preferred lapse $N$, the second the angular height offset
\begin{equation}\label{eq:angularheight}
 \Delta h(r,z)\equiv h(r,z)-h(r,0)=u(r,z)-u(r,0),
\end{equation}
and the third the scalar squared shear $\sigmasq$.

The height offset gives a direct, but coordinate-dependent, picture of
how the preferred leaves depart angularly from the equatorial reference
within the prescribed stationary chart.  Indeed, with the
horizon-regular reference time $t_{\rm ref}=t_{\rm BL}+h_{\rm ref}(r)$,
our physical preferred time satisfies $t=T=t_{\rm ref}+u(r,z)$; a
constant-$T$ leaf is therefore represented by
$t_{\rm ref}=T-u(r,z)$.  Subtracting its equatorial value removes the
entire radial reference height and makes $\Delta h$ finite across the
future horizon.  Unlike this coordinate height, $\sigmasq$ is a
scalar invariant of the preferred foliation.  These maps are not
isometric embeddings of the intrinsic three-geometry; they display
three complementary fields on the numerical $(r,z)$ domain.

The maps reveal that the computed solutions have nontrivial angular
structure while retaining a positive lapse across the future
black-hole horizon and into the sampled trapped region.  They also
make visually explicit that similar local shear patterns on the two
branches do not imply an identical VCDM state: the four-geometries,
asymptotic conditions, and lapse fields differ.  All mapped values
are reconstructed from the archived spectral states, without any new
nonlinear solve.  Independent four-dimensional reconstructions at
$3948$ sampled points per branch give
$\max|K-K_0|=3.11\times10^{-10}$ (AF) and
$3.32\times10^{-11}$ (dS) on the displayed sampling grid.  No
continuation below the inner excision radius is inferred.

\subsection{What the lapse-zero diagnostics establish}
To bridge our analytical results from Sec.~\ref{sec:zero} with the numerical solutions, we verified the exact algebraic lapse--flux identity~\eqref{eq:barrier} across all seventeen archived configurations using an independent post-processing implementation. Across the seventeen runs, this audit evaluates more than $5.6 \times 10^5$ points in the inner domain---explicitly including the black-hole horizon $r_+$, the cosmological horizon $r_c$, and the limiting expressions on the rotation axis $z = 1$---as well as dense off-grid samples across the compactified outer domain. The maximum normalized discrepancy between the two sides of Eq.~\eqref{eq:barrier} over the entire dataset is $1.3 \times 10^{-15}$, confirming agreement at approximately double-precision roundoff level. At every sampled point, the regularized lapse denominator $\mathcal D$ in Eq.~\eqref{eq:D} is strictly positive. (As noted earlier, we report these diagnostics as dense finite-grid verifications rather than formal interval-arithmetic proofs.)

\begin{figure}[tbp]
 \centering
 \includegraphics[width=\columnwidth]{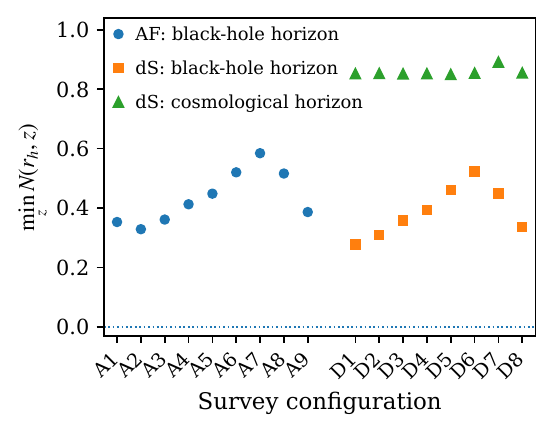}
 \caption{Minimum lapse on a 129-point angular grid at the black-hole horizon for all seventeen archived solutions; the cosmological-horizon minima are also shown for the eight dS cases. Labels refer to Table~\ref{tab:S8}. Every point lies strictly above zero, and the exact horizon relation \eqref{eq:horizonN} independently ties these margins to nonzero normal flux; its future orientation is checked separately. These minima are sampled diagnostics, not rigorous all-angle bounds.}\label{fig:horizonmargin}
\end{figure}

Figure~\ref{fig:horizonmargin} displays the minimum value of the lapse function across a 129-point angular grid on the future black-hole horizon $r_+$ for all seventeen solutions, together with the minimum lapse on the future cosmological horizon $r_c$ for the eight de Sitter configurations. Over the entire survey, the smallest black-hole horizon lapse encountered is $\min_z N_+ = 0.27675$, which sits comfortably above zero. Moreover, every single configuration satisfies the sharper sufficient angular-fluctuation bounds~\eqref{eq:Csufficient} on the horizon grid, meaning that the angle-averaged flux margin $|C - \kappa r_h J_h|$ strictly dominates the maximum angular fluctuation $\|\delta P_h\|_\infty$. For the two principal rotating benchmarks ($a = 0.6, q = 0.4, \kappa = 0.1, C = 2$), the horizon lapse minima are
\begin{align}
 \min_z N_+&\simeq0.44837 &&(\Le=0),\\
 \min_z N_+&\simeq0.39307 &&(\Le=0.03),\\
 \min_z N_c&\simeq0.85199 &&(\Le=0.03).
\end{align}
By virtue of the exact horizon relation~\eqref{eq:horizonN}, these strictly positive lapse margins imply nonzero $|P|$. The separately verified signs show strictly inward normal flux ($P > 0$) on $r_+$ and strictly outward normal flux ($P < 0$) on $r_c$ at every sampled angle. Thus, none of the computed rotating configurations is near a horizon lapse breakdown on the tested horizon grids.

It is equally important to state clearly what our interior numerical domain does---and does not---establish inside the black hole. In every configuration listed in Table~\ref{tab:S8}, the solved domain penetrates into the trapped region below $r_+$ by a radial depth of at least $r_+ - r_{\rm in} \ge 0.1928M$, and the lapse $N$ remains strictly positive down to $r = r_{\rm in}$. We do \emph{not} find or claim a numerical root $N = 0$ at smaller radii $r < r_{\rm in}$ simply because we do not integrate the PDE below the excision boundary $r_{\rm in}$. Rather, the logical division of labor between our numerics and our analytics is the following: the numerical solutions demonstrate that a smooth family of VCDM-compatible CMC slices crosses the future horizons and enters the trapped region $r < r_+$ without pathology, while the exact analytical bounds of Eqs.~\eqref{eq:zeroDelta}--\eqref{eq:graphineq} guarantee that if an inward continuation of these slices eventually encounters a stationary spacelike lapse-zero limiting surface, that surface can only lie inside the trapped block $r_- < r < r_+$. Determining whether a given exterior solution terminates at a simple zero or asymptotes to a rotating trumpet below $r_{\rm in}$ requires a dedicated interior continuation study.

\subsection{Foliation dependence at fixed spacetime}\label{sec:shear_results}
\begin{figure*}[t]
\centering
\includegraphics[width=0.47\textwidth]{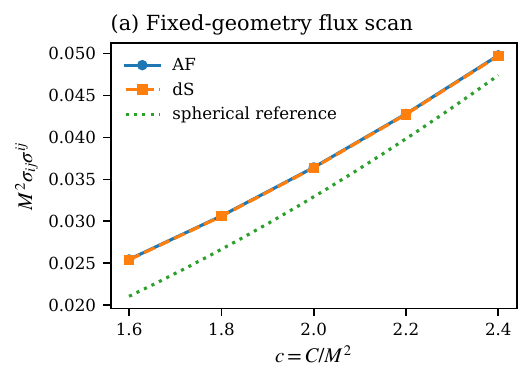}\hfill
\includegraphics[width=0.47\textwidth]{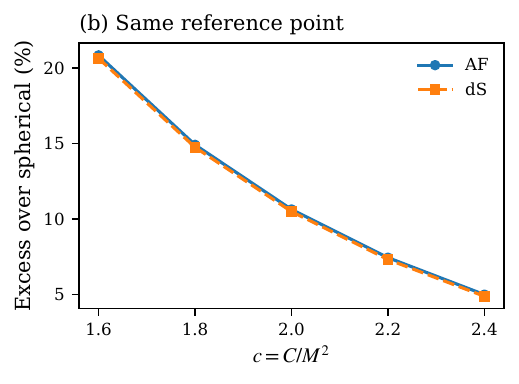}
\caption{Dependence of the preferred-foliation shear on $c=C/M^2$,
with $a/M=0.6$, $q/M=0.4$, $\kappa M=0.1$, $r_{\rm in}/M=1.5$, and
$g=0$. Each branch keeps its own four-geometry fixed. (a) $s=M^2\sigmasq$
at $(r/M,z)=(3,0.5)$; markers denote the five computed configurations,
connecting segments are guides over the measured interval, and the
dotted reference is the exact spherical result. (b) Fractional excess
above that reference. Connecting segments in (b) are guides only.
No continuation to $C=0$ or beyond the displayed interval is implied.}
\label{fig:Cscan}
\end{figure*}

Because a VCDM configuration is specified by both the four-metric and the preferred slicing, it is essential to exhibit a coordinate-independent geometric scalar that changes when the preferred foliation is varied on a \emph{fixed} four-dimensional spacetime. A natural diagnostic is the squared tracefree extrinsic curvature (squared shear) of the preferred slices:
\begin{equation}\label{eq:shear}
 \sigmasq=K_{ij}K^{ij}-\frac{K_0^2}{3}.
\end{equation}
Holding $(M,a,q,\Le,K_0)$ fixed while varying $C$ changes the
preferred normal congruence but leaves the four-metric, electromagnetic
field, and metric horizons unchanged. We now extend the two-point
comparison in the original survey to
\[
 c\equiv C/M^2\in\{1.6,1.8,2.0,2.2,2.4\},
\]
with $a/M=0.6$, $q/M=0.4$, $\kappa M=0.1$, and
$r_{\rm in}/M=1.5$. The zero-mean inner profile is kept at $g(z)=0$;
the mean inner flux changes with $C$ according to Eq.~\eqref{eq:innerbc}.
Within each branch the background is exactly fixed: $\Le=0$ for AF and
$\Le M^2=0.03$ for dS. Define $s=M^2\sigmasq$, evaluated at
$(r/M,z)=(3,0.5)$. The measured values are given in
Table~\ref{tab:Cscan} and Fig.~\ref{fig:Cscan}.

\begin{table}[tbp]
\caption{Fixed-four-geometry flux scan, at $(r/M,z)=(3,0.5)$.
The other parameters and the inner prescription are fixed as stated in
the text. The last column is the exact spherical reference
$s_{\rm sph}=6c^2/3^6$, not a fit to the rotating solutions.
Displayed digits identify the numerical results and are not uniform
accuracy claims.}
\label{tab:Cscan}
\begin{ruledtabular}\begin{tabular}{cccc}
$c$ & $s_{\rm AF}$ & $s_{\rm dS}$ & $s_{\rm sph}$\\
1.6 & 0.02546297 & 0.02542142 & 0.02106996\\
1.8 & 0.03064462 & 0.03060088 & 0.02666667\\
2.0 & 0.03642268 & 0.03637640 & 0.03292181\\
2.2 & 0.04279755 & 0.04274842 & 0.03983539\\
2.4 & 0.04976959 & 0.04971732 & 0.04740741\\

\end{tabular}\end{ruledtabular}
\end{table}

In this interval the shear increases with $c$ in both branches, while its
fractional excess above the spherical reference decreases. For AF the
excess falls from $20.85\%$ at $c=1.6$ to $4.98\%$ at $c=2.4$; for dS
it falls from $20.65\%$ to $4.87\%$. The absolute AF/dS difference at
matched $c$ is approximately $(4.16$--$5.23)\times10^{-5}$, below $0.2\%$
of the shear. These comparisons describe this point and this selected
boundary-value family, not a universal monotonicity theorem.

A quadratic fit is a useful numerical description on
$1.6\le c\le2.4$:
\begin{align}\label{eq:Cfits}
 s_{\rm AF}(c)&\simeq0.00746001c^2+0.00054306c+0.00549653,
 \nonumber\\
 s_{\rm dS}(c)&\simeq0.00745602c^2+0.00054560c+0.00546114.
\end{align}
The maximum relative residuals of the five-point quadratic fits are
$4.9\times10^{-6}$ and $5.4\times10^{-6}$, respectively. Thus they are
not exact quadratic laws. In particular, the constant terms cannot be
interpreted as the shear at $C=0$, where this regular family has not been
continued. A centered slope estimate about $c=2$ gives
$\dd s/\dd c\simeq0.03038$ (AF) and $0.03037$ (dS), compared with
$0.03292$ for the spherical formula. The corresponding local logarithmic
slope is approximately $1.67$, not the spherical value $2$.

All ten configurations satisfy the sampled numerical gates. Repeating
$c=1.6,2.0,2.4$ on finer grids gives maximum relative shear changes
$7.93\times10^{-14}$ (AF) and $2.68\times10^{-12}$ (dS) on nine common
near-hole probes. This is a stability test, not an error bound on all
points: in AF the prolonged state already satisfies the stopping
tolerance, whereas dS takes an additional Newton step. The numerical
scope is detailed in Appendix~\ref{app:newscans}. The physical conclusion
is the invariant distinction between preferred foliations on the same
four-geometry, not a new metric multipole or a test-matter signature.

\begin{figure}[tbp]
 \centering
  \includegraphics[width=\columnwidth]{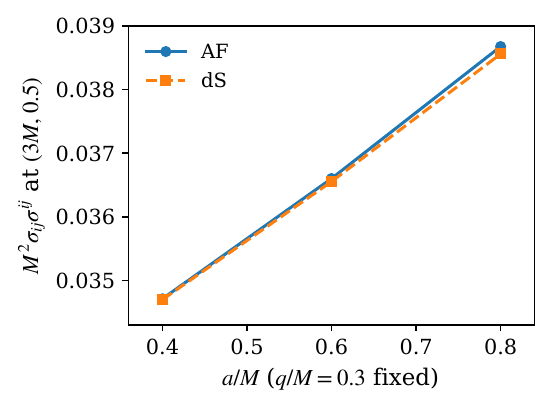}
 \caption{Preferred-slice squared shear at $r=3M$, $z=0.5$, for fixed $q/M=0.3$, $\kappa M=0.1$, and $C/M^2=2$. Each marker is a separately solved nonlinear boundary-value problem. The higher-spin case uses $r_{\rm in}=1.3M$ rather than $1.5M$; the curve therefore describes the selected family, not a derivative at completely fixed inner boundary data.}\label{fig:spin}
\end{figure}

We also examined how the sampled squared shear $\sigmasq(3M, 0.5)$ responds to changes in black-hole spin $a$ and electric charge $q$ across the survey:
\begin{itemize}
 \item Holding $q/M = 0.3$, $\kappa M = 0.1$, and $C/M^2 = 2$ fixed, increasing the spin from $a/M = 0.4$ through $0.6$ to $0.8$ monotonically increases $\sigmasq(3M, 0.5)$ on both the asymptotically flat and de Sitter branches, as shown in Fig.~\ref{fig:spin}.
 \item Holding $a/M = 0.6$, $\kappa M = 0.1$, and $C/M^2 = 2$ fixed, increasing the charge across $q/M = 0.3, 0.4, 0.6$ produces a mild decrease in $\sigmasq(3M, 0.5)$.
\end{itemize}
We note that these trends describe the behavior of our specific boundary-value family at the fiducial point $(r, z) = (3M, 0.5)$---and that the highest-spin ($a/M = 0.8$) and highest-charge ($q/M = 0.6$) runs impose the inner boundary condition~\eqref{eq:innerbc} at $r_{\rm in} = 1.3M$ rather than $1.5M$ in order to stay inside $r_+$. They should therefore be read as concrete properties of the computed configurations rather than universal monotonicity or stability theorems.

Finally, on the high-resolution de Sitter benchmark, our independent four-dimensional metric and Christoffel reconstruction finds finite $K_{ij}K^{ij}$ consistent with smooth crossing at the sampled points on both the future black-hole horizon $r_+$ and the future cosmological horizon $r_c$. As a further geometric check of the reconstructed slices, we verified that the null Killing horizon generators $\partial_t + (a/J_h)\partial_\psi$ have vanishing norm at $r = r_h$, and that the proper two-dimensional area of each constant-$t$ horizon cross section $r = r_h$ agrees to numerical precision with the standard Kerr--Newman--de Sitter horizon area $4\pi(r_h^2 + a^2)/\Xi$. (Geometrically, this area is independent of the residual height profile $u(r_h, z)$ because $\Dr(r_h) = 0$ eliminates the $e^2$ term in the regular metric~\eqref{eq:regularmetric}, so the induced two-metric on the slice $t = \text{const}$, $r = r_h$ has determinant $\det\gamma_{AB} = (r_h^2 + a^2)^2/\Xi^2$ in $(z, \psi)$ coordinates regardless of $u_z$.) Together, these four-dimensional checks confirm the geometric fidelity of the reconstructed preferred foliations.

\subsection{A supplementary inner-profile test at fixed \texorpdfstring{$C$}{C}}\label{sec:boundarytest}
The boundary freedom in Sec.~\ref{sec:innerdata} can be tested without
changing the four-geometry or the mean curvature. Independently of the
seventeen-case survey, we solved both benchmark branches with
$M=1$, $a=0.6$, $q=0.4$, $\kappa=0.1$, $C=2$, and
$r_{\rm in}=1.5$, replacing the inner data by
\eqref{eq:quadrupolebc} with $\epsilon/M^2=0,\pm0.02$.
The original solver equations and outer prescriptions were unchanged;
only the additive inner-flux profile was varied.

Each choice was recomputed at two resolutions: $(84,32,16)$ and
$(100,40,20)$ for AF, and $(88,40,16)$ and $(104,48,20)$ for dS.
Here the entries count inner radial, outer radial, and angular nodes,
respectively. The matching radii remained $48M$ and $24M$.
Table~\ref{tab:innerprofile} gives a preferred-foliation scalar at a
fixed exterior point from the finer grids.
\begin{table}[tbp]
 \caption{Supplementary fixed-$C$ boundary-profile test.
 Values of $M^2\sigmasq$ at $(r/M,z)=(3,0.5)$ for the two
 branches; all other parameters are specified in the text.
 These new tests are separate from the frozen seventeen-case survey.}
 \label{tab:innerprofile}
 \begin{ruledtabular}\begin{tabular}{ccc}
 $\epsilon/M^2$ & AF & dS\\
 $-0.02$ & $0.03643457849$ & $0.03638478554$\\
 $0$     & $0.03642268258$ & $0.03637640402$\\
 $+0.02$ & $0.03641082473$ & $0.03636803951$\\
 \end{tabular}\end{ruledtabular}
\end{table}

Across these twelve solves, the maximum sampled inner CMC errors
were below $3\times10^{-11}$ and $5\times10^{-12}$ in the AF and
dS branches, respectively; the sampled outer residuals remained below
$6\times10^{-10}$ and $10^{-6}$. The original survey-level flux,
matching, and horizon-sign criteria were also satisfied. On a common
$59\times21$ comparison grid over $1.8\le r/M\le8$, the medium/fine
lapse differences were below $5\times10^{-14}$ (AF) and
$10^{-14}$ (dS). The changes relative to $\epsilon=0$ were instead
approximately $2\times10^{-3}$ in the largest sampled lapse
difference. The exterior shear change in Table~\ref{tab:innerprofile}
provides an invariant distinction between the foliations, not merely
a change in the time label.

These results give numerical examples of additional solutions of
\emph{different inner-boundary problems} at the same $C$ and with the
same leading asymptotic conditions. They do not prove solvability for
an arbitrary function $g$, nonuniqueness for fixed complete data, or
regular extension below the cutoff. The supplementary computations
were performed during this revision and are supplied separately for
independent reproduction; the archived seventeen states were not
modified.

\subsection{Paired AF/dS comparison at decreasing mean curvature}
\label{sec:smallk}
The smallness of cosmological curvature and the proximity of the
preferred foliation to BL are distinct questions. To test the first
without resetting the inner angular data, we compare pairs with
\begin{equation}\label{eq:pairedchoices}
\begin{array}{ll}
\mathrm{AF}:&\Le=0,\quad K_0=3\kappa,\\
\mathrm{dS}:&\Le=3\kappa^2,\quad K_0=3\kappa,
\end{array}
\end{equation}
for $\varepsilon\equiv\kappa M=0.10,0.05,0.03,0.01$. In each pair we
keep $a/M=0.6$, $q/M=0.4$, $C/M^2=2$, $r_{\rm in}/M=1.5$, and $g=0$.
The same numerical value of $\kappa$ is used on both branches. Across
the sequence the prescribed mean flux changes with $\kappa$ through
Eq.~\eqref{eq:innerbc}; it is the zero-mean profile, not the entire inner
flux, that is held fixed. Matching radii are numerical interfaces and
need not coincide between branches.

The comparison identifies the displayed metric parameters and BL-derived
coordinates $(r,z)$, not a de Sitter conserved mass with an AF ADM mass.
It also retains the common convention $\partial_\tau t=1$ for the
stationary time. As explained in Sec.~\ref{sec:asymptoticstate}, realizing
both members of each pair in a specified VCDM potential requires
compatible auxiliary states and is not guaranteed by fixing $\kappa$.
The scan is consequently a comparison of stationary geometric
boundary-value problems with VCDM embeddings, not a simulated
cosmological history in one prescribed model.

There is an exact spherical reference. At equal $(M,q,C,\kappa)$,
Eq.~\eqref{eq:sphericals} gives
\begin{equation}
\begin{split}
 N^2={}&1-\frac{2(M+\kappa C)}r+\frac{q^2}{r^2}+\frac{C^2}{r^4}\\
       &+\left(\kappa^2-\frac{\Le}{3}\right)r^2.
\end{split}
\end{equation}
\begin{align}\label{eq:sphericalpair}
 N_{\rm AF}^2-N_{\rm dS}^2&=\kappa^2r^2,\\
 (\sigmasq)_{\rm AF}-(\sigmasq)_{\rm dS}&=0,\\
 n^r_{\rm AF}-n^r_{\rm dS}&=0.
\end{align}
The last two identities use spherical symmetry and do not follow in the
rotating case. In particular, rotation can change the coefficient of the
lapse difference without changing its observed power of $\varepsilon$.

\begin{figure*}[t]
\centering
\includegraphics[width=0.32\textwidth]{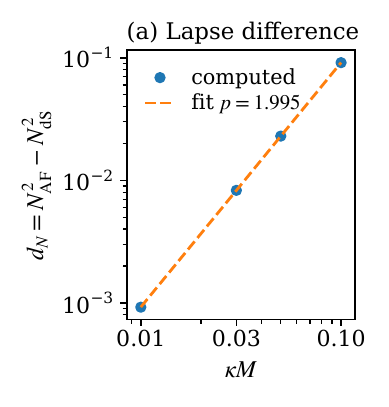}\hfill
\includegraphics[width=0.32\textwidth]{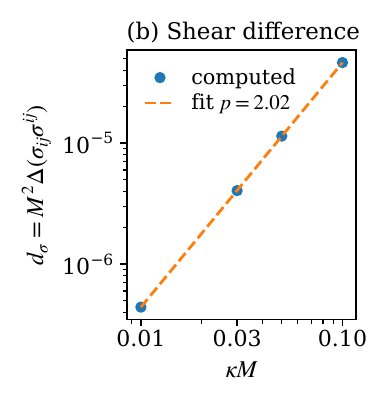}\hfill
\includegraphics[width=0.32\textwidth]{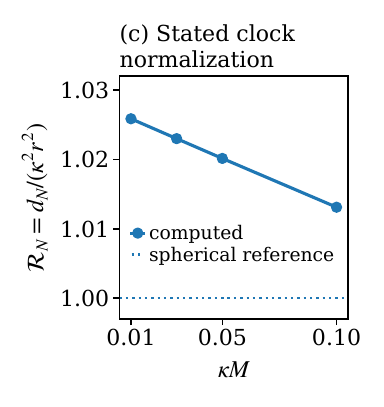}
\caption{Paired AF/dS local comparison for equal $\kappa$ within each pair
and the fixed data specified in Sec.~\ref{sec:smallk}. (a) Squared-lapse
difference $d_N$ in the numerical time normalization. (b) Difference
$d_\sigma$ of the preferred shear scalars. Lines in (a,b) are descriptive
power-law fits restricted to $0.01\le\kappa M\le0.10$.
(c) $\mathcal R_N=d_N/(\kappa^2r^2)$, with the exact spherical reference
$\mathcal R_N=1$; the segments joining measured points are guides.
No curve is extrapolated to zero. The ratio in (c) is not a
clock-normalization-independent rotational observable. Input precision
is stated in Table~\ref{tab:smallk}.}
\label{fig:smallk}
\end{figure*}

At the common exterior point $(r/M,z)=(3,0.5)$ define
\begin{align}\label{eq:pairedobservables}
 d_N&\equiv N_{\rm AF}^2-N_{\rm dS}^2,\\
 d_\sigma&\equiv M^2\big[(\sigmasq)_{\rm AF}
                            -(\sigmasq)_{\rm dS}\big],\\
 \mathcal R_N&\equiv d_N/(\kappa^2r^2).
\end{align}
Table~\ref{tab:smallk} and Fig.~\ref{fig:smallk} show that both positive
differences decrease strongly along the sequence. Descriptive,
unweighted log--log fits to these four points give
\begin{equation}\label{eq:smallkfits}
 d_N\propto\varepsilon^{1.995},\qquad
 d_\sigma\propto\varepsilon^{2.02}
 \quad(0.01\le\varepsilon\le0.10).
\end{equation}
These are finite-range effective slopes, not proofs of a leading
$O(\varepsilon^2)$ expansion or statistical measurements. The final
state-derived exports give slopes $1.99478565$ and $2.02152530$;
Eq.~\eqref{eq:smallkfits} deliberately retains only the precision warranted
for a descriptive four-point comparison. Seventeen-digit floating-point
exports identify the stored numerical values, not their continuum error.

\begin{table}[tbp]
\caption{Paired local comparison at $(r/M,z)=(3,0.5)$, with the parameters
and normalization of Sec.~\ref{sec:smallk}. Values are displayed at
limited precision; state-derived full-precision exports are retained
in the reproducibility archive. The listed digits are not numerical
uncertainty estimates.}
\label{tab:smallk}
\setlength{\tabcolsep}{3pt}
\begin{ruledtabular}\begin{tabular}{cccc}
$\varepsilon$ & $d_N$ & $d_\sigma$ & $\mathcal R_N$\\
0.10 & $9.117934\times10^{-2}$ & $4.6279\times10^{-5}$ & 1.01310\\
0.05 & $2.295334\times10^{-2}$ & $1.1426\times10^{-5}$ & 1.02015\\
0.03 & $8.286339\times10^{-3}$ & $4.0504\times10^{-6}$ & 1.02300\\
0.01 & $9.232871\times10^{-4}$ & $4.4082\times10^{-7}$ & 1.02587\\

\end{tabular}\end{ruledtabular}
\end{table}

The ratio $\mathcal R_N$ rises from $1.01310$ to $1.02587$ over the
computed range. We do not infer an exact limit, or a universal
``rotation correction'', from an extrapolation to $\varepsilon=0$.
This caution is also required by clock normalization. Replacing the dS
time by its asymptotically cosmological normalization in
Eq.~\eqref{eq:cosmicclock}, while retaining the AF convention, gives
\begin{equation}\label{eq:lapserescalingpair}
 \widetilde d_N=d_N-(\Xi^2-1)N_{\rm dS}^2,
 \qquad \Xi=1+\kappa^2a^2.
\end{equation}
The change is itself of order $\kappa^2$ at fixed $a$ and finite lapse.
For example, $\widetilde d_N/(\kappa^2r^2)\simeq0.99369$ at
$\varepsilon=0.01$ in this alternative convention. Thus the coefficient
of $d_N$ is meaningful only with its time normalization stated.
The shear of a given foliation is unchanged by such a clock relabeling,
although comparing its values across two different metrics still uses
the specified identification of points.

At $\varepsilon=0.01$, the final AF values at $(r/M,z)=(3,0.5)$ are
\begin{align*}
 N_{\rm AF}&=0.63502560225938,\\
 M^2(\sigmasq)_{\rm AF}&=0.03536139941298.
\end{align*}
The displayed dS values are $0.634298217288$ and $0.035360958591$,
respectively. Two-grid tests and separate
continuous-residual checks support these local comparisons, with the
scope and arithmetic specified in Appendix~\ref{app:newscans}. In
particular, an excellent agreement of local observables did not suffice
while the asymptotic residual failed its gate; only the accepted final
states are used here.

The small shear difference should be compared with the variation of
each shear between accepted representations, not with its PDE residual
alone. At $\varepsilon=0.01$, let $s_X=M^2(\sigmasq)_X$ at the fiducial
point and let $e_X$ be the maximum reported relative shear change on
the corresponding common comparison grid. That grid includes this
point. The conservative sampled envelopes in Table~\ref{tab:signalgrid}
give
\begin{equation}\label{eq:signalsamplingratio}
 \frac{e_{\rm AF}\max|s_{\rm AF}|+e_{\rm dS}\max|s_{\rm dS}|}
 {|d_\sigma|}<1.0\times10^{-4},
\end{equation}
where each maximum in the numerator is over the two accepted
representations at this point. This compares the signal with observed
grid sensitivity. It is not a bound on the shared systematic error or
a statistical significance.

\begin{table}[tbp]
\caption{Grid sensitivity relative to the shear signal at
$\kappa M=0.01$, $(r/M,z)=(3,0.5)$. The upper envelopes use the
reported maxima of relative differences on common grids, rather than
claiming pointwise continuum error bars. Values are rounded upwards.}
\label{tab:signalgrid}
\begin{ruledtabular}\begin{tabular}{lcc}
Branch & $e_X$ & $e_X\max|s_X|$\\
AF & $5.21\times10^{-10}$ & $1.85\times10^{-11}$\\
dS & $6.59\times10^{-10}$ & $2.34\times10^{-11}$\\
\end{tabular}\end{ruledtabular}
\end{table}

This evidence supports the local similarity of the selected AF and dS
configurations over the computed finite range. It is not uniform in
radius: at $r\sim\kappa^{-1}$ their asymptotic structures differ, with
$N_{\rm AF}\sim\kappa r$ and $N_{\rm dS}\to\Xi^{-1}$.
Nor does it construct matching to matter-filled FLRW, determine the
normal tilt relative to BL for all admissible inner data, or fix the
value of $C$ selected by collapse.

\section{Physical interpretation as VCDM black holes}\label{sec:discussion}
The configurations constructed above solve a stationary VCDM
boundary-value problem on the stated domain. Their physical content
is the metric together with $T=t$, the auxiliary state,
and the foliation boundary data. We now distinguish that exact
stationary result from a conditional approximation to black holes in
an evolving universe, and from the separate dynamical problem that
would select their inner data.

\subsection{Connection to an evolving cosmological state}
\label{sec:cosmomatch}
In spatially flat FLRW VCDM~\cite{DeFeliceDollMukohyama2020}, let a dot denote differentiation with
respect to proper cosmological time $t_{\rm cos}$, rather than an
unconverted derivative with respect to the numerical coordinate $t$.
Then
\begin{align}\label{eq:cosmobg}
 3H_{\rm cos}&=\frac32 V_{,\varphi}-\varphi,&
 \dot\varphi&=\frac{3(\rho+p)}{2M_{\rm Pl}^2},\nonumber\\
 3H_{\rm cos}^2&=\frac{\rho}{M_{\rm Pl}^2}+\Lambda_\varphi,&
 \Lambda_\varphi&=V-\varphi V_{,\varphi}
                   +\frac34V_{,\varphi}^{\,2}.
\end{align}
Here $\rho,p$ denote ordinary cosmological matter, not the
black-hole electromagnetic field, and $M_{\rm Pl}$ is the reduced
Planck mass. These equations also underlie sign-changing effective
dark energy in $\Lambda_{\rm s}$VCDM
\cite{AkarsuDeFeliceEtAl2026}. They imply
\begin{align}\label{eq:cosmoidentities}
 V_{,\varphi\varphi}
 &=\frac23+\frac{4M_{\rm Pl}^2\dot H_{\rm cos}}{3(\rho+p)},\nonumber\\
 &=\frac{2M_{\rm Pl}^2}{9(\rho+p)}
        \frac{\dd\Lambda_\varphi}{\dd\ln a_{\rm cos}},\\
 \dot\Lambda_\varphi
 &=6H_{\rm cos}\dot H_{\rm cos}
       +\frac{3H_{\rm cos}(\rho+p)}{M_{\rm Pl}^2}.\nonumber
\end{align}
The divided expressions require $\rho+p\ne0$; using $\ln a_{\rm cos}$
as a time variable also requires $H_{\rm cos}\ne0$. These identities
connect potential curvature to the actual expansion history, rather
than treating $V_{,\varphi\varphi}$ as an unrelated large parameter.

At an epoch $t_*$, freezing $\varphi_*=\varphi(t_*)$ in
Eqs.~\eqref{eq:K0fixed}--\eqref{eq:fixedpointLambda} gives
\begin{equation}\label{eq:instantmatch}
 K_*=3H_{\rm cos}(t_*),\qquad \Le(\varphi_*)=\Lambda_\varphi(t_*).
\end{equation}
This is an exact match of the \emph{algebraic auxiliary data} in the
same theory. It is not an exact matter-filled cosmological metric.
Indeed,
\begin{equation}\label{eq:mattermismatch}
 \kappa_*^2-\frac{\Le(\varphi_*)}{3}
       =\frac{\rho(t_*)}{3M_{\rm Pl}^2}.
\end{equation}
Our computed $\kappa=H_\Lambda$ family realizes the vacuum de Sitter
relation. An epoch with matter generally requires independent
$\kappa$ and $\Le$ in the frozen electrovacuum reference. The flat
branch $\Le=0$, $\kappa>0$ can satisfy the instantaneous algebraic
data when $\Lambda_\varphi(t_*)=0$, without making the full FLRW
universe asymptotically flat.

No smallness assumption on the absolute $\varphi_*$ or $V(\varphi_*)$
is required. For example, the field-origin change
\begin{align}\label{eq:fieldorigin}
 \widetilde\varphi&=\varphi+c,\\
 \widetilde V(\widetilde\varphi)
 &=V(\widetilde\varphi-c)+\frac{2c}{3}\widetilde\varphi-\frac{c^2}{3}
 \nonumber
\end{align}
preserves $3V_{,\varphi}/2-\varphi$, $\Lambda_\varphi$, and
$V_{,\varphi\varphi}$, at corresponding field values. This is a
simultaneous field and potential redefinition, not freedom to change
the state while holding an arbitrary potential function fixed.

\subsection{Assumptions for an instantaneous local description}
\label{sec:localapprox}
A stationary reference is a candidate leading-order description in a
near zone $r=O(M)$ if cosmological curvature and its variation are
small compared with the black-hole scales. A useful hierarchy is
\begin{equation}\label{eq:cosmosmall}
 \begin{gathered}
 |H_{\rm cos}|M\ll1,\quad |\Le|M^2\ll1,\quad
 M^2|\dot H_{\rm cos}|\ll1,\\
 M^2|\dot\varphi|\ll1,\quad M^3|\dot\Lambda_\varphi|\ll1,\quad
 \frac{M^2(|\rho|+|p|)}{M_{\rm Pl}^2}\ll1.
 \end{gathered}
\end{equation}
One also needs a region in which the near-zone solution can be
matched to the cosmological spacetime, slow evolution of the hole
and its boundary data on the scale $M$, and a regular solution of the
elliptic constraints with those data, without parametrically enhanced
response to the neglected cosmological terms. These regularity and
matching properties are assumptions, not consequences of scale
separation alone. We assume no additional local
matter source or incoming disturbance that invalidates the frozen
electrovacuum approximation. Under these assumptions, the exact
stationary solutions provide local reference black holes for an
adiabatic matching calculation. We have not constructed that matched
expansion or bounded its error. In particular, the numerical
benchmarks, including the extension down to $\kappa M=0.01$, do not
constitute simulations at astrophysical scale separations. The paired
scan of Sec.~\ref{sec:smallk} tests local differences between stationary
references, not the error of a matched FLRW construction.

The geometric correction to Kerr--Newman can be small without making
the preferred slicing close to BL. Equation~\eqref{eq:BLboost} measures
the relative normal tilt within each geometry. For the spherical
reference, wherever $f>0$,
\begin{equation}\label{eq:sphericaltiltlimit}
 \tanh^2\eta=\frac{(C/r^2-\kappa r)^2}
                    {f+(C/r^2-\kappa r)^2}.
\end{equation}
As $\kappa M\to0$ and $\Le M^2\to0$ at fixed $r/M$ and fixed nonzero
$C/M^2$, the $C/r^2$ term need not vanish. Zero mean curvature therefore
does not imply the BL foliation. At a horizon crossed with finite
positive $N$, the divergent rapidity relative to BL reflects degeneration
of the BL reference normal, not a singularity of the regular preferred
normal. These facts explain why a small local AF/dS difference does not
eliminate the independent foliation data. Collapse and global matching
must select those data; neither their unique selection nor a memory of
a particular formation history has been established here.

\subsection{Why the frozen approximation need not reach a lapse zero}
\label{sec:nonuniform}
Slow cosmological evolution does not automatically imply slow local
evolution along the preferred normal near $N=0$. Suppose the
auxiliary field in the matched solution remains homogeneous on the
leaves, $\varphi=\varphi(t)$ in unitary gauge. Since $n^\mu\nabla_\mu T=1/N$,
\begin{align}\label{eq:normalrates}
 n^\mu\nabla_\mu\varphi&=\frac{\partial_t\varphi}{N},&
 \nabla_\mu\varphi\nabla^\mu\varphi
 &=-\frac{(\partial_t\varphi)^2}{N^2}.
\end{align}
Likewise, if $K=K(t)$, then
$n^\mu\nabla_\mu K=(\partial_t K)/N$. Here $N$ and $\partial_t$ refer to the same preferred-time
normalization. In the de Sitter reference of
Eq.~\eqref{eq:cosmicclock}, $\partial_t\varphi=\dot\varphi/\Xi$
and $N_{\rm cos}=\Xi N$, so the same rate is
$\partial_t\varphi/N=\dot\varphi/N_{\rm cos}$. More general
cosmological matching must likewise specify the time conversion. On the
open domain, a relabeling $\widetilde T=f(T)$ divides both
$\partial_T\varphi$ and $N$ by $f'$, leaving their ratio unchanged.
Thus rescaling the time label cannot remove a divergent normal rate;
the same-leaf obstruction in Sec.~\ref{sec:clockregularity} also prevents
curing a localized lapse zero while retaining a regular exterior clock.

Thus $M^2|\partial_t\varphi|/N$ and
$M^2|\partial_t K|/N$ are additional local diagnostics of the
frozen approximation. On a region with a lapse bounded away from
zero they can be small, whereas approaching $N=0$ can amplify the
same background rates. If $\partial_t\varphi$ remains nonzero as
$N\to0$, the scalar in Eq.~\eqref{eq:normalrates} diverges. A regular
evolving solution must avoid this combination, suppress the
numerator appropriately, or cease to follow the frozen reference.
This kinematic observation identifies a possible nonuniform inner
region; it is not a computed instability or a solution of its
dynamics.

Consequently, the critical spherical estimate
$R_*/r_{\rm h}\simeq3/4$ in Sec.~\ref{sec:sphericalzero} describes a
particular stationary reference, not the measured endpoint of a
cosmological rotating black hole. Its infinite proper distance does
not supply a dynamical selection rule. The finite-time or
infinite-time approach to an endpoint, and the associated validity
of the effective description, require evolution. The safe-time-domain
result in the spherical collapse model of
Ref.~\cite{DeFeliceMaedaMukohyamaPookkillath2023} cannot simply be
transferred to a general rotating solution with $\dot\varphi\ne0$.

\subsection{What standard matter can and cannot distinguish}
\label{sec:matterprobes}
Consider a diffeomorphism-invariant test-matter action of the form
\begin{align}\label{eq:matteraction}
 S_{\rm m}&=S_{\rm m}[g_{\mu\nu},\mathsf A_\mu,\psi],\nonumber\\
 \frac{\delta S_{\rm m}}{\delta T}
 &=\frac{\delta S_{\rm m}}{\delta\varphi}
 =\frac{\delta S_{\rm m}}{\delta\lambda}
 =\frac{\delta S_{\rm m}}{\delta\lambda_2}=0.
\end{align}
where the equalities state the absence of direct preferred-field
couplings at fixed metric and electromagnetic field. Standard minimal
coupling is a sufficient example. Two stationary configurations with
the same $g_{\mu\nu}$ and $F_{\mu\nu}$ then give identical test-matter
equations, when physical probe states and boundary conditions are
also identified. For standard minimally coupled probes, neutral
particles and geometric-optics light rays follow the same geodesics;
charged particles follow the same Lorentz-force law. Changing $C$ or the inner angular flux does not
therefore create a direct background signature in such probes.
The local foliation shear can change without changing a metric
multipole or this test-matter response.

Diffeomorphism invariance \emph{alone} is not sufficient for the
conclusion. A term depending on the scalar
$n^\mu\nabla_\mu\psi$, for example, can be covariant and yet couple
matter directly to the preferred foliation. Such couplings are not
assumed here. Conversely, even minimally coupled matter can probe
the theory indirectly through metric backreaction, gravitational
perturbations, or formation dynamics. We establish background
indistinguishability under the stated conditions, not an equivalence
of all VCDM and GR observables. No prediction for ringdown or
cosmological-transition signatures is extracted from the stationary
solutions.

\subsection{Collapse selection and sign-changing cosmologies}
\label{sec:collapsechoice}
The data to be selected by formation include the integrated flux
$C$ and the angular inner profile $g(z)=\delta P(r_{\rm in},z)$, as
well as the possible interior limiting behavior. Once a global
solution is obtained, its restriction to any chosen cutoff fixes
these data there; they are not arbitrary adjustable inputs to that
same solution. Whether distinct initial conditions yield identical
late-time data is a dynamical question. The quadrupolar test proves
that different inner boundary problems can yield different exterior
preferred normals, not that collapse leaves arbitrary independent
hair. Moving $r_{\rm in}$ while retaining the same numerical profile
need not preserve the physical solution.

The placement of the cutoff inside the metric horizon does not make
the elliptic stationary problem insensitive to it. Nor does an
elliptic equation, by itself, prove instantaneous physical signaling:
GR initial data also obey constraints often solved elliptically.
The full constrained VCDM evolution must determine the physical role
of inner data. No cosmic-censorship or universal late-time-selection
result is inferred here.

In $\Lambda_{\rm s}$VCDM, $\Lambda_\varphi<0$ can coexist with
$H_{\rm cos}^2>0$ because of the matter contribution to
Eq.~\eqref{eq:cosmobg}. The full universe is then not anti-de Sitter.
A frozen electrovacuum reference with $\Le<0$ would instead use a
Kerr--Newman--AdS metric with different global boundary data; that
stationary construction is not part of the present numerics.
For small $|\Le|M^2$, either sign gives a small local metric
correction, subject to the matching assumptions above.

A smooth transition of amplitude $|\Delta\Lambda_\varphi|=O(H_*^2)$
over proper time $\tau_{\rm tr}$, with $|H_{\rm cos}|$ remaining of order
$H_*>0$ and bounded away from zero throughout the transition, gives the estimate
\begin{equation}\label{eq:transitionparameter}
 M^2|\dot H_{\rm cos}|
 =O\!\left[(MH_*)\frac{M}{\tau_{\rm tr}}\right]
   +O\!\left[\frac{M^2|\rho+p|}{M_{\rm Pl}^2}\right].
\end{equation}
A transition rapid on a Hubble scale can still be slow on the scale
$M$. If $\tau_{\rm tr}\lesssim M$, an adiabatic sequence is no longer
justified by time-scale separation, even though the absolute metric
correction may remain small. On FLRW,
\begin{equation}\label{eq:lambdarate}
 \dot\lambda=-V_{,\varphi\varphi}\dot\varphi
            =-2\dot H_{\rm cos}-\frac23\dot\varphi.
\end{equation}
The preferred constraint sector may therefore respond differently
from the background metric. Near a small lapse, normal rates can be
further amplified as in Eq.~\eqref{eq:normalrates}. A burst,
instability, or destruction of a horizon is not a consequence of a
sign change alone. Smooth transition models avoid the singular
background derivatives of an ideal discontinuous switch
\cite{AkarsuDeFeliceEtAl2026}; their black-hole response requires the
time-dependent problem.

\section{Conclusions}\label{sec:conclusions}
We have constructed stationary rotating VCDM configurations on the
constant-auxiliary Einstein--Maxwell branch, specifying the preferred
foliation as well as the Kerr--Newman or Kerr--Newman--de Sitter metric.
Their nonzero-CMC leaves cross the future black-hole horizon and, in
dS, the future cosmological horizon. The horizon-regular flux
formulation and separate asymptotic prescriptions provide the
boundary-value construction beyond the general GR embedding criterion.

The seventeen-case survey and supplementary scans demonstrate regular
configurations for multiple metric and foliation data. At fixed
four-geometry, changing $C$ or the zero-mean inner flux profile changes
the shear of the preferred normal. The five-point $C$ scan quantifies
this distinction; it does not generate a new metric multipole or a
direct signature for standard test matter. Paired AF/dS calculations
down to $\kappa M=0.01$ show finite-range local differences with
effective powers close to two. The lapse coefficient depends on time
normalization, while the shear is invariant under relabeling of each
foliation. Neither comparison determines a universal zero-curvature
limit or erases the different asymptotic ends.

The exact lapse identity excludes $N=0$ in the regular $\Dr>0$
stationary exterior, and nonzero oriented horizon fluxes ensure a
positive horizon lapse in the constructed configurations. The BL
normal is genuinely singular at a simple horizon, as shown by its
acceleration. More generally, a lapse zero approached along a leaf
cannot be removed by time relabeling while keeping the lapse finite and
positive on a regular portion of that leaf. For the stationary ansatz,
$N=-n_\mu\xi^\mu$ with the prescribed Killing normalization. This
nonremovability does not equate every endpoint with a local singularity:
a zero reached along a finite-proper-length path forces unbounded normal
acceleration, whereas a trumpet can be an asymptotic end at infinite
proper distance with finite normal invariants. The conditional placement
bounds do not locate or classify the endpoint of the uncomputed rotating
interior.

Continuous residuals, geometric reconstructions, and selected
cross-language and grid comparisons support the numerical
boundary-value solutions. They are finite-sample checks, with their
scope stated in the appendices. The stationary auxiliary data can
match a cosmological state algebraically, but a local adiabatic
interpretation also needs scale separation and compatible elliptic
and matching data. Collapse selection, a full FLRW matching
construction, and gravitational perturbations remain distinct
problems. The present result supplies explicit, regular preferred
strong-field reference configurations for those investigations.

\section*{Data availability and computational disclosure}
The numerical states, source codes, diagnostics, full-precision observable
exports, and figure scripts have been assembled in a versioned research
archive and are available from the author on reasonable request.

Maple, NumPy, SciPy, SymPy, mpmath, Matplotlib, and Julia were used. ChatGPT assisted
with derivations, software, diagnostics, figures, and drafting.
Scientific verification, authorship, and responsibility remain with
the author.

\appendix
\section{Asymptotic coefficients and their assumptions}\label{app:asymptotics}
In this appendix we record the second-order asymptotic coefficient $u_2(z)$ on the matched expanding de Sitter branch ($\kappa = \HL > 0$) and explain how it is uniquely determined by the compactified CMC equation~\eqref{eq:compact}.

Defining the combined mass--flux scale $B \equiv M + \kappa C$ and the dimensionless rotational parameter $A \equiv \kappa^2 a^2$ (so that $\Dz = 1 + A z^2$ and $\Xi = 1 + A$), the quadratic coefficient $u_2(z)$ in the asymptotic expansion~\eqref{eq:dsasym} reads
\begin{align}\label{eq:u2}
 u_2(z)=\frac{\kappa a^2}{4}\bigg[&\frac{3B^2(1+\Xi z^2)}{\Dz}
 -\frac{\Xi q^2z^2}{\Dz}\nonumber\\
 &-\frac{\Xi q^2z\operatorname{arsinh}(\kappa a z)}
 {\kappa a\Dz^{3/2}}\bigg].
\end{align}
Notice that $u_2(z)$ is manifestly even in $z$ (since $z\operatorname{arsinh}(\kappa a z)$ is the product of two odd functions), is completely smooth along the rotation axis $z = \pm 1$, and vanishes identically in the spherically symmetric limit $a \to 0$ (where $\operatorname{arsinh}(\kappa a z)/(\kappa a) \to z$ and the prefactor $\kappa a^2/4 \to 0$).

We emphasize that Eq.~\eqref{eq:u2} is an exact analytical consequence of expanding the normal-flux conservation law~\eqref{eq:compact} in powers of $\rho = 1/r$ around $\rho = 0$, rather than a numerical fit. Specifically, when the power-series ansatz $u(\rho, z) = u_{\rm dS}(z) + \rho u_1(z) + \rho^2 u_2(z) + O(\rho^3)$ is substituted into the subtracted fluxes $\Pi$ and $\Theta$ in Eq.~\eqref{eq:regflux}, the compactified equation $-\partial_\rho\Pi + \partial_z\Theta = 0$ at the first nontrivial order in $\rho$ reduces to an inhomogeneous second-order linear ordinary differential equation in $z$ for $v(z) \equiv u_2(z)$. The linear differential operator acting on $v(z)$ is the Legendre-like operator deformed by the oblate parameter $A = \kappa^2 a^2$:
\begin{align}\label{eq:angularop}
 \mathcal L_A v={}&\partial_z[(1-z^2)\Dz v']
 +3Az(1-z^2)v'\nonumber\\
 &+2(1+2A-4Az^2)v,
\end{align}
while the source is fixed by the lower-order coefficients and the
parameters. The uniqueness of a smooth even solution is not inferred
merely from the $A=0$ limit. For every $A\ge0$, an exact odd
homogeneous solution is
\begin{equation}
 y(z)=\frac{z}{(1+Az^2)^{3/2}}.
\end{equation}
Multiplication of $\mathcal L_A$ by $\Dz^{3/2}$ puts its derivative
part in self-adjoint form with
$p(z)=(1-z^2)\Dz^{5/2}$. For any homogeneous solution $v$,
$p(yv'-y'v)$ is constant. If $v$ is regular at $z=1$, this constant
vanishes because $p(1)=0$. At $z=0$, $y(0)=0$, $y'(0)=1$, and
$p(0)=1$, so $v(0)=0$. Even parity also gives $v'(0)=0$; uniqueness of
the ordinary initial-value problem then implies $v\equiv0$.
Thus there is no regular even homogeneous ambiguity. At $A=0$, the
regular zero mode is the \emph{odd} Legendre polynomial
$\mathsf P_1(z)=z$, consistently excluded by parity.

Direct substitution of \eqref{eq:u2} into the expanded equation
verifies the particular solution. The argument fixes this coefficient
on the assumed smooth power-series branch; it does not exclude
solutions with different asymptotic regularity or logarithmic terms.

For completeness, let us also record the leading $\rho \to 0$ behavior of the subtracted fluxes~\eqref{eq:regflux} on the de Sitter branch. Using $u_{\rm dS}(z)$ and $u_1(z)$ from Eq.~\eqref{eq:u1}, the subtracted polar flux satisfies $\QQ - \QQ_{\rm dS} = O(\rho^2)$ (so that $\Theta = (\QQ - \QQ_{\rm dS})/\rho^2$ has a finite limit at $\rho = 0$), while the subtracted radial flux approaches the anisotropic limiting profile
\begin{equation}
 \Pi(0,z)=C+\frac{B}{\kappa}
 \left(\frac{\sqrt\Xi}{\Dz^{3/2}}-1\right).
\end{equation}
Using the elementary integral $\int_0^1 \Dz^{-3/2}\,\dd z = [z/\sqrt{1 + \kappa^2 a^2 z^2}]_0^1 = \Xi^{-1/2}$, one verifies immediately that the second term in $\Pi(0, z)$ has zero angular mean, so $\int_0^1 \Pi(0, z)\,\dd z = C$ in exact agreement with the radial first integral~\eqref{eq:firstint}. Finally, note that the asymptotically flat hyperboloidal expansion~\eqref{eq:afasym} (where $\HL = 0$ is fixed while $\kappa > 0$) has a different radial scaling of the lapse ($N \sim \kappa r$ rather than $N \to \Xi^{-1}$) and therefore cannot be obtained by taking $\kappa \to 0$ inside the matched de Sitter formulas~\eqref{eq:u1} and \eqref{eq:u2}.

\section{Spherical double-root and regularity checks}\label{app:spherical}
In this appendix we provide the explicit algebraic relations governing the double-root trumpet configurations in the exact spherically symmetric limit ($a = 0$) discussed in Sec.~\ref{sec:sphericalzero}, together with an independent check of the elliptic lapse equation~\eqref{eq:lapseU}.

Expanding the squared spherical lapse $N^2 = S(r) = f(r) + \beta(r)^2$ from Eq.~\eqref{eq:sphericals} in powers of $r$ yields
\begin{equation}
 S(r)=1-\frac{2(M+\kappa C)}r+\frac{q^2}{r^2}
   +\frac{C^2}{r^4}+(\kappa^2-\HL^2)r^2.
\end{equation}
A double root at $r = R_*$ satisfies the simultaneous polynomial conditions $S(R_*) = 0$ and $S'(R_*) = 0$. Taking the linear combination $S(R_*) + R_* S'(R_*) = 0$ eliminates the $M + \kappa C$ term immediately and solves for $C_*^2$ in terms of $R_*$; substituting $C_*^2$ back into $S(R_*) = 0$ then gives $M + \kappa C_*$ in terms of $R_*$:
\begin{align}\label{eq:doubleroot}
 C_*^2&=\frac{R_*^4-q^2R_*^2}{3}
            +(\kappa^2-\HL^2)R_*^6,\\
 M+\kappa C_*&=\frac{2R_*^2+q^2}{3R_*}
            +(\kappa^2-\HL^2)R_*^3.
\end{align}
Note that because squaring $\beta(R_*) = C_*/R_*^2 - \kappa R_*$ loses track of the sign of $\beta(R_*)$, the algebraic system~\eqref{eq:doubleroot} by itself includes spurious roots that do not belong to the future-oriented black-hole branch. To isolate the physical future trumpet end, one must supplement Eqs.~\eqref{eq:doubleroot} with the future inward-flux condition $\beta(R_*) > 0$ (which gives Eq.~\eqref{eq:Ccurve}), placement inside the trapped region $r_- < R_* < r_+$, and the local minimum condition $S''(R_*) > 0$.

The exact spherical solution also provides a transparent analytical verification of the stationary CMC elliptic lapse equation~\eqref{eq:lapseU}. On the constant-$t$ spherical slice with induced three-metric $\dd\ell^2 = S(r)^{-1}\dd r^2 + r^2\dd\Omega^2$ (for which $\sqrt{\gamma} = r^2 S^{-1/2}\sin\theta$ and $\gamma^{rr} = S$), the spatial Laplacian acting on the radial function $N(r) = \sqrt{S(r)}$ evaluates to
\begin{align*}
 D^2 N &=\frac{\sqrt S}{r^2}\frac{\dd}{\dd r}
       \left(r^2\sqrt S\,\frac{\dd\sqrt S}{\dd r}\right)\\
 &=\frac{\sqrt S}{r^2}\frac{\dd}{\dd r}\left(\frac{r^2 S'}2\right)
 =N\left(\frac{S'}r+\frac{S''}2\right).
\end{align*}
Dividing by $N$ and differentiating $S(r)$ term by term---noting that the $1/r$ term $-2(M + \kappa C)/r$ is harmonic in three-dimensional flat space and therefore drops out of $S'/r + S''/2$---we obtain
\begin{equation}
 \frac{D^2N}{N}=\frac{S'}r+\frac{S''}2
 =3(\kappa^2-\HL^2)+\frac{6C^2}{r^6}+\frac{q^2}{r^4}.
\end{equation}
Comparing this expression with the general potential $\mathcal U = \sigmasq + 3(\kappa^2 - \HL^2) + E_i E^i + B_i B^i$ in Eq.~\eqref{eq:lapseU}, we see that $6C^2/r^6$ matches the exact spherical squared shear $\sigmasq$ in Eq.~\eqref{eq:sphericalK}, $q^2/r^4$ matches the squared electric field $E_i E^i = q^2/r^4$ (with $B_i = 0$), and $3(\kappa^2 - \HL^2) = K_0^2/3 - \Le$ matches the cosmological shift. This provides an independent check of both the relative signs and the electromagnetic normalization in Eq.~\eqref{eq:lapseU}.

As emphasized in Sec.~\ref{sec:sphericalzero}, comparing Eq.~\eqref{eq:sphericalK} with the squared four-acceleration~\eqref{eq:accel}, $a_i a^i = [S'(r)]^2/[4S(r)]$, demonstrates that finiteness of $K_{ij}K^{ij}$ and $\sigmasq$ is necessary but not sufficient for regularity of the preferred normal congruence: at a simple zero of $S(r)$, $K_{ij}K^{ij}$ remains finite while $a_i a^i$ diverges, whereas at a double root $R_*$ both invariants remain finite. Generalizing double-root trumpet conditions to rotating maximal or CMC slicings requires a dedicated two-dimensional free-boundary analysis inside the trapped region (see, e.g., Ref.~\cite{MauryaGuttiNigam2024} for rotating maximal limiting surfaces) and cannot be inferred merely by setting $\mathcal F(r) = 0$ in the spherical formulas above.

\section{Survey data and reproducibility}\label{app:survey}

\paragraph{Cross-language reproduction.}
The Julia full-grid runs on the author's machine used
$(84,36,16)$ nodes for AF and $(88,40,16)$ for dS. The respective
sampled inner CMC residuals were $3.03\times10^{-11}$ and
$2.76\times10^{-12}$, the outer regularized residuals
$1.80\times10^{-10}$ and $2.72\times10^{-7}$, and mean-flux errors
$1.54\times10^{-11}$ and $2.22\times10^{-11}$.
At $(r,z)=(3M,0.5)$, the lapse differences from the frozen Python
references were $3.33\times10^{-16}$ in both branches, with shear
squared differences $5.53\times10^{-14}$ and $1.39\times10^{-14}$.
These are pointwise and sampled comparisons, not uniform error
bounds. All seventeen frozen states were evaluated in Julia, but
this log does not establish seventeen fresh nonlinear Julia solves.
The six quadrupolar runs reproduce the local observables in
Table~\ref{tab:innerprofile}; their displayed log is not a second
full off-grid audit of every perturbed state.

Table~\ref{tab:S8} summarizes the seventeen archived numerical configurations, labeled A1--A9 on the asymptotically flat hyperboloidal branch ($\Le = 0$) and D1--D8 on the expanding de Sitter branch ($\Le = 3\kappa^2$); these same labels are used on the horizontal axis of Fig.~\ref{fig:horizonmargin}. All geometric diagnostics and residuals quoted in Table~\ref{tab:S8} are taken directly from the frozen archived dataset, while independent audit reruns by the author on Linux reproduce all acceptance criteria within expected floating-point roundoff variations in the last digits and residual plateaus. Across the entire archived survey, the largest off-grid inner CMC residual is $\delta K = \max|K - K_0| \simeq 5.6 \times 10^{-9}$ (occurring in configuration D1 on the de Sitter branch, with even smaller inner residuals across the AF branch).

\begin{table*}[t]
 \caption{The seventeen archived survey configurations, in units $M=1$.
 AF means $\Le=0$ and dS means $\Le=3\kappa^2$.
 $N_3$ and $\sigma_3^2$ are evaluated at $(r,z)=(3,0.5)$.
 The final column gives sampled inner CMC and outer compactified
 residual maxima from the archived runs, not uniform error bounds.
 The supplementary inner-profile experiment is not included in this table.}
 \label{tab:S8}
 \setlength{\tabcolsep}{4pt}\renewcommand{\arraystretch}{1.06}
 \begin{ruledtabular}\begin{tabular}{llccccccc}
 ID & Branch & $a$ & $q$ & $\kappa$ & $C$ & $N_3$ & $\sigma_3^2$ & $\delta K\ /\ \mathcal R_{\rm comp}$\\
 A1 & AF & 0.0 & 0.4 & 0.10 & 2.0 & 0.5976 & 0.03292 & $\scriptstyle 6.1\times10^{-16} / 1.3\times10^{-14}$ \\
A2 & AF & 0.4 & 0.0 & 0.10 & 2.0 & 0.5905 & 0.03482 & $\scriptstyle 4.2\times10^{-9} / 1.6\times10^{-10}$ \\
A3 & AF & 0.4 & 0.3 & 0.10 & 2.0 & 0.5987 & 0.03471 & $\scriptstyle 1.1\times10^{-9} / 9.5\times10^{-11}$ \\
A4 & AF & 0.6 & 0.3 & 0.10 & 2.0 & 0.6077 & 0.03660 & $\scriptstyle 1.1\times10^{-10} / 1.0\times10^{-10}$ \\
A5 & AF & 0.6 & 0.4 & 0.10 & 2.0 & 0.6138 & 0.03642 & $\scriptstyle 2.6\times10^{-11} / 2.1\times10^{-10}$ \\
A6 & AF & 0.8 & 0.3 & 0.10 & 2.0 & 0.6194 & 0.03868 & $\scriptstyle 5.2\times10^{-10} / 2.8\times10^{-10}$ \\
A7 & AF & 0.6 & 0.6 & 0.10 & 2.0 & 0.6309 & 0.03591 & $\scriptstyle 1.2\times10^{-10} / 2.1\times10^{-10}$ \\
A8 & AF & 0.6 & 0.4 & 0.06 & 2.0 & 0.6086 & 0.03594 & $\scriptstyle 7.3\times10^{-12} / 5.8\times10^{-10}$ \\
A9 & AF & 0.6 & 0.4 & 0.10 & 1.8 & 0.6172 & 0.03064 & $\scriptstyle 3.9\times10^{-11} / 5.0\times10^{-11}$ \\
D1 & dS & 0.4 & 0.0 & 0.10 & 2.0 & 0.5083 & 0.03481 & $\scriptstyle 5.6\times10^{-9} / 1.6\times10^{-6}$ \\
D2 & dS & 0.4 & 0.3 & 0.10 & 2.0 & 0.5178 & 0.03470 & $\scriptstyle 2.3\times10^{-10} / 9.2\times10^{-8}$ \\
D3 & dS & 0.6 & 0.3 & 0.10 & 2.0 & 0.5275 & 0.03655 & $\scriptstyle 6.7\times10^{-12} / 1.1\times10^{-6}$ \\
D4 & dS & 0.6 & 0.4 & 0.10 & 2.0 & 0.5344 & 0.03638 & $\scriptstyle 1.8\times10^{-12} / 9.5\times10^{-7}$ \\
D5 & dS & 0.8 & 0.3 & 0.10 & 2.0 & 0.5397 & 0.03856 & $\scriptstyle 1.1\times10^{-11} / 5.0\times10^{-7}$ \\
D6 & dS & 0.6 & 0.6 & 0.10 & 2.0 & 0.5537 & 0.03587 & $\scriptstyle 1.1\times10^{-11} / 5.2\times10^{-7}$ \\
D7 & dS & 0.6 & 0.4 & 0.08 & 2.0 & 0.5580 & 0.03615 & $\scriptstyle 2.9\times10^{-13} / 1.1\times10^{-7}$ \\
D8 & dS & 0.6 & 0.4 & 0.10 & 1.8 & 0.5382 & 0.03060 & $\scriptstyle 2.4\times10^{-12} / 2.1\times10^{-7}$ \\

 \end{tabular}\end{ruledtabular}
\end{table*}

The spectral collocation grids employ $(N_r^{\rm in}, N_\rho^{\rm out}, N_z) = (84, 36, 16)$ nodes for the asymptotically flat runs and $(88, 40, 16)$ nodes for the de Sitter runs, with radial matching interfaces placed at $R_m = 48M$ (AF) and $R_m = 24M$ (dS). Configurations with moderate spin and charge use an inner excision radius $r_{\rm in} = 1.5M$, whereas the high-spin ($a = 0.8M$) and high-charge ($q = 0.6M$) configurations use $r_{\rm in} = 1.3M$ so that $r_{\rm in}$ remains comfortably inside the outer horizon $r_+$.

To be accepted into the validated survey, each configuration is required to pass the following automated quality-control gates:
\begin{enumerate}
 \item convergence of the coupled two-domain Newton--Raphson solver;
 \item off-grid inner mean-curvature error $|K - K_0| < 5 \times 10^{-8}$ across 500 deterministic off-grid sample points in $(r, z) \in [r_{\rm in}, R_m] \times [0, 1]$;
 \item off-grid outer regularized compactified residual $|\mathcal R_{\rm comp}| < 2 \times 10^{-7}$ (AF) or $< 5 \times 10^{-6}$ (dS) across 600 deterministic off-grid sample points in $(\rho, z) \in [0, 1/R_m] \times [0, 1]$;
 \item conservation of the angle-integrated radial flux~\eqref{eq:firstint} and interface continuity at $r = R_m$ to better than $2 \times 10^{-8}$;
 \item correct future horizon flux orientations and positive sampled
lapse denominators, together with positive spatial-metric eigenvalues
at the independent geometric probe; the denser geometric benchmark
audits are separate from this per-case survey check; and
 \item an independent four-dimensional Christoffel-symbol reconstruction check at the fiducial point $(r, z) = (3M, 0.5)$ confirming $|K - K_0| < 10^{-7}$.
\end{enumerate}

The horizon-margin and lapse-identity verification scripts read the frozen spectral coefficient arrays directly without modifying them. They evaluate the solution on a 129-point angular grid along each horizon ($r_+$ and $r_c$), on a dense logarithmic radial grid in the inner domain augmented by the exact horizon radii, and across the compactified outer domain. All source codes and archived data arrays are verified via cryptographic SHA-256 hashes, allowing the exact algebraic identities, numerical horizon margins, and spherical reference roots to be regenerated via a single automated verification command.

Finally, to ensure that the validation suite actually catches defective configurations rather than passing unconditionally, the package includes explicit \emph{negative controls}: test cases with a positive mean flux $\bar P(r_+) > 0$ that violates the pointwise condition $P(r_+, z) > 0$ at some angles, incompatible auxiliary-field potential values violating Eqs.~\eqref{eq:potential}--\eqref{eq:lambdaeff}, an intentionally misnormalized Maxwell stress-energy tensor from early development tests, and the artificial pointwise isotropic outer boundary condition $\Pi(0, z) = C$ at $\IR$ on the AF branch. All negative controls are properly flagged and rejected by the diagnostic suite. While our independent multiprecision evaluations, Christoffel reconstructions, and alternative interpolation checks provide strong internal cross-verification, we note that a fully independent discretization (e.g., finite elements or shooting in a different coordinate gauge) remains a natural avenue for future numerical comparisons.

\section{Supplementary flux and small-curvature scans}
\label{app:newscans}
This appendix records the tests added after the original seventeen-case
survey. All numerical residuals below use $M=1$; dimensionless forms are
$M|K-K_0|$, $|\mathcal R_{\rm comp}|/M^3$, and the mean-flux error divided
by $M^2$. A ``sampled pass'' means satisfaction of the specified tests on
the stated finite sets. It is not a uniform error estimate, a continuum
existence theorem, a stability analysis, or a uniqueness result.

\subsection{Fixed-geometry scan in the mean flux}
The ten $C$-scan configurations use $(n_I,n_O,n_z)=(84,36,16)$ for AF
and $(88,40,16)$ for dS, with $R_m/M=48$ and $24$, respectively.
The physical data are those of Sec.~\ref{sec:shear_results}. Each branch
starts from $C/M^2=2$ and uses neighboring accepted states as continuation
seeds. The maximum reported compactified residuals are
$3.96\times10^{-10}$ (AF) and $1.97\times10^{-7}$ (dS); all ten pass the
survey-level checks, including the inner CMC and mean-flux tests.
The archived $C/M^2=1.8,2.0$ values provide additional local regression
checks.

For $C/M^2=1.6,2.0,2.4$, the finer grids are $(104,44,18)$ (AF) and
$(108,48,18)$ (dS), with unchanged matching radii and Taylor orders.
Comparisons use nine common near-hole probes. The maximum relative
changes are $7.93\times10^{-14}$ and $2.68\times10^{-12}$ for the shear,
and $5.28\times10^{-16}$ and $2.83\times10^{-14}$ for the lapse.
The fine-grid compactified residuals remain below $4.07\times10^{-10}$
and $5.28\times10^{-7}$. The AF prolongations already satisfy the Newton
tolerance at iteration zero; each dS case takes one additional update.
These tests establish sampled stability under the stated prolongation
and refinement, not an independently seeded solve or a known convergence
order. A three-point curvature and centered-slope comparison at
$C/M^2=1.6,2.0,2.4$ is stable under this refinement. It does not turn the
five-point quadratic fit into an exact law.

\begin{table*}[t]
\caption{Final small-$\kappa$ checks at $\kappa M=0.01$. All values are in
units $M=1$. Entries $n_I,n_M,n_F,n_z$ count grid nodes; AF has only one
outer domain, denoted $n_O$. Residuals are maxima on the respective
reported sample sets, not supremum bounds. The dS seam column uses the
separate continuous one-sided audit; its sampling differs from the
solver's extra seam rows.}
\label{tab:smallkaudit}
\setlength{\tabcolsep}{5pt}
\begin{ruledtabular}\begin{tabular}{llccccc}
Branch & Grid & Nodal PDE & $\max|K-K_0|$ & Outer $|\mathcal R_{\rm comp}|$
& Mean-flux error & Far seam $|\mathcal R_{\rm comp}|$\\
AF & $(64,24,12)$ & $1.40\times10^{-9}$ & $1.13\times10^{-9}$
 & $2.65\times10^{-8}$ & $6.91\times10^{-11}$ & ---\\
AF & $(80,32,16)$ & $1.56\times10^{-10}$ & $1.42\times10^{-11}$
 & $3.77\times10^{-8}$ & $1.64\times10^{-11}$ & ---\\
dS & $(144,44,48,18)$ & $4.52\times10^{-10}$ & $4.71\times10^{-11}$
 & $1.05\times10^{-7}$ & $1.95\times10^{-14}$ & $6.63\times10^{-7}$\\
dS & $(144,44,64,18)$ & $1.17\times10^{-9}$ & $1.41\times10^{-10}$
 & $4.51\times10^{-10}$ & $7.11\times10^{-15}$ & $1.43\times10^{-9}$\\
\end{tabular}\end{ruledtabular}
\end{table*}

\subsection{Direct AF flux evaluation at \texorpdfstring{$\kappa M=0.01$}{kappa M=0.01}}
Finite-order Taylor evaluation of the exterior fluxes must be distinguished
from the asymptotic boundary condition itself. At small $\kappa$, the AF
coefficients in Eq.~\eqref{eq:afasym} contain inverse powers of $\kappa$,
and increasing spatial resolution does not remove finite-order radial
series error. The final AF implementation retains the same unknown $w$
and the same natural boundary row at $\rho=0$, but evaluates the
untruncated finite-$\rho$ algebraic fluxes directly in multiprecision.
A short formal series is used only to extract the regular endpoint
limit. The exact oblate Minkowski height was examined diagnostically,
but no new reference-height subtraction is required in the accepted runs.

Two Python-generated candidates with $R_m/M=160$ are reevaluated and
updated in Julia at 320-bit flux precision. The stored states, spectral
matrices, and Newton linear algebra are still in double precision; this
is not a full arbitrary-precision boundary-value solve. Local tests
compare the fluxes with multiprecision reference values, check the local
analytic Jacobian, and reproduce the exact spherical limit. A coupled
Jacobian directional test has relative discrepancy $1.17\times10^{-9}$.
Continuous off-grid checks use 459 inner probes and 70 outer probes,
supplemented by horizon, mean-flux, interface, and geometric tests.
This targeted exterior probe set differs from the 600-point survey set;
its maxima should not be interpreted as a denser uniform audit or as
supremum bounds. The independent direct-flux evaluation and two-grid
comparison are complementary checks, not a replacement for coverage of
all radii and angles.
The coarser candidate needs no Julia Newton update; the finer takes one,
reducing the nodal residual from $3.18\times10^{-9}$ to
$1.56\times10^{-10}$. Both retain the nodal tolerance $3\times10^{-9}$
and the original AF physical residual thresholds.

\subsection{De-aliased constrained dS solve at \texorpdfstring{$\kappa M=0.01$}{kappa M=0.01}}
The final dS calculation uses one inner radial domain and two compactified
outer domains, with $R_m/M=192$ and $r_{\rm split}/M=768$. The inner domain
has 144 radial nodes, the middle 44, and the far domain 48 or 64; both
states use 18 angular nodes. At the added interface, continuity of $v$
and $\Pi$ is equivalent to continuity of $u$ and $P$ because the same
analytic subtraction is used on both sides. The original asymptotic
Dirichlet data $v(0,z)=u_2(z)$ are unchanged.

Fluxes are evaluated on oversampled grids before their derivatives are
restricted to the solution nodes, using radial factor $1.5$ and angular
factor $2$. This reduces the interpolation/aliasing defect identified
by comparisons with continuous flux derivatives. Let $F_{\rm PDE}$
collect the remaining de-aliased PDE rows, $q$ the far-side seam PDE,
and $B$ the physical boundary and transmission equations. The final
iteration minimizes
\begin{equation}\label{eq:constrainedobjective}
 \frac12\|F_{\rm PDE}(x)\|_2^2+
 \frac{\lambda_{\rm num}^2}{2}\|q(x)\|_2^2,
 \qquad B(x)=0,
\end{equation}
with $\lambda_{\rm num}=10$ and equality-constraint tolerance
$5\times10^{-11}$. Here $\lambda_{\rm num}$ is a numerical weight,
not the VCDM auxiliary field $\lambda$. The extra seam residual is not
a third physical interface condition. Both the discrete residual and
the continuous one-sided PDE must satisfy their separate tests.

The finer solve uses a thin singular-value decomposition for the reduced
regularized least-squares step, rather than forming its normal matrix.
This avoids squaring the ill-conditioning of the reduced map. The
48-node reference used the earlier Schur implementation, whereas the
64-node result uses the SVD step and is warm-started from that reference.
Their comparison therefore tests agreement of two accepted final
representations with different linear-algebra implementations, not a
pure refinement experiment with an otherwise identical algorithm.
The reported termination and the finite iteration budget do not by
themselves establish a vanishing optimization gradient. Acceptance is
instead based on the residual and constraint tests in
Table~\ref{tab:smallkaudit} and the independent one-sided interface audit.

\subsection{Scope of agreement and retained limitations}
The dS equality-constraint maxima are $2.51\times10^{-11}$ and
$1.87\times10^{-11}$; the new interface flux jump is below $10^{-15}$.
The independent continuous radial-flux derivative jump decreases from
$6.63\times10^{-7}$ to $1.31\times10^{-9}$, and the final middle-domain
residual stays below $3.3\times10^{-10}$. The dS validation requests
1200 off-grid samples and includes extra edge/layer probes. Its
one-sided audit is separate from the optimization but shares some
flux and interpolation infrastructure, so is not a fully independent
discretization.

On common comparison points, the maximum relative AF changes are
$3.45\times10^{-12}$ in the near-hole lapse and $5.21\times10^{-10}$
in the shear; for dS they are $3.23\times10^{-14}$ and
$6.58\times10^{-10}$. These are differences between numerical
representations, not certified continuum accuracy. The outer AF
residual is not monotone between grids, and the dS refinement changes
principally the far radial domain. All retained configurations pass
the sampled positivity, horizon-sign, and flux checks. Full-precision observables were subsequently reconstructed from the
frozen states by the common export pipeline and retained with the
validation logs. The finite-range fits do not assign statistical error
bars or infer a zero-curvature limit.

\end{document}